\documentclass[traditabstract]{aa}  
\usepackage{graphicx}
\usepackage{txfonts}
\newcommand{\kep}{{\em Kepler}}

\newcommand{\teff}{\ensuremath{T_{\mathrm{eff}}}}
\newcommand{\logg}{\ensuremath{\log g}}

\newcommand{\pdot}{$\dot{\it p}$}
\newcommand{\msun}{${\mathrm{M}}_{\odot}$} 
\newcommand{\rsun}{${\mathrm{R}}_{\odot}$} 
\newcommand{\lsun}{${\mathrm{L}}_{\odot}$} 
\newcommand{\muHz}{$\mu$Hz}

\newcommand{\mjup}{$\rm M_{\rm Jup}$}
\newcommand{\bjdtdb}{BJD$_{\rm TDB}$}
\def\apjl{ApJ}%
\def\aap{A\&A}%
\usepackage{color}

\begin{document} 

\title{The sdB pulsating star V391~Peg and its putative giant planet
revisited after 13 years of time-series photometric data
\thanks{The complete set of data is available in electronic form
at the CDS via anonymous ftp to cdsarc.u-strasbg.fr (130.79.128.5)
or via http://cdsweb.u-strasbg.fr/cgi-bin/qcat?J/A+A/}
\thanks{Based on observations obtained at the following observatories:
WHT 4.2m, TNG 3.6m, Calar Alto 2.2m, NOT 2.5m, Loiano 1.5m, LOAO 1.0m, 
MDM 1.3m, Moletai 1.6m, MONET-North 1.2m, Piszk\'{e}stet\H{o} 1.0m, 
Mercator 1.2m, Wise 1.0m, Lulin 1.0m, Baker 0.6m.}
%ESO Paranal Observatory (programme 075.D-0371).}
}

\subtitle{}

\author{R. Silvotti\inst{1}, S. Schuh\inst{2}, S.-L. Kim\inst{3}, R. 
Lutz\inst{4}, M. Reed\inst{5}, S. Benatti\inst{6}, R. Janulis\inst{7},\\
L. Lanteri\inst{1}, R. \O stensen\inst{5}, T.~R. Marsh\inst{8}, V.~S. 
Dhillon\inst{9,10},
%C.~D. Savoury\inst{10}, 
M. Paparo\inst{11}, L. Molnar\inst{11}
%, W.~S. Hsiao\inst{12}
% \and
% C. Ptolemy\inst{2}\fnmsep\thanks{Just to show the usage
% of the elements in the author field}
}

\institute{INAF--Osservatorio Astrofisico di Torino, strada dell'Osservatorio
20, 10025 Pino Torinese, Italy\\
\email{silvotti@oato.inaf.it}
%\email{silvotti@oato.inaf.it,\,lanteri@oato.inaf.it}
%\and
%Institut f\"ur Astrophysik, Georg-August-Universit\"at G\"ottingen, 
%Friedrich-Hund-Platz 1, 37077 G\"ottingen, Germany
%\\
%\email{schuh@astro.physik.uni-goettingen.de}
\and
Max Planck Institute for Solar System Research, Justus-von-Liebig-Weg 3, 
37077 G\"ottingen, Germany
%\email{schuh@mps.mpg.de}
%\and
%Institut f\"ur Astronomie und Astrophysik, Kepler Center for Astro and 
%Particle Physics, Eberhard-Karls-Universit\"at, Sand 1, 72076 T\"ubingen, 
%Germany
\and
Korea Astronomy and Space Science Institute, Daejeon 34055, Korea
%\email{slkim@kasi.re.kr}
\and
German Aerospace Center (DLR), Remote Sensing Technology Institute, 
M\"unchener Str. 20, 82234 We\ss ling, Germany
%\email{Ronny.Lutz@dlr.de}
\and
Department of Physics, Astronomy and Materials Science, Missouri State 
University, Springfield, MO 65897, USA
%\email{mikereed@missouristate.edu,\,roy@ostensen.eu}
\and
INAF-Osservatorio Astronomico di Padova, Vicolo dell'Osservatorio 5, 35122 
Padova, Italy
%\email{serena.benatti@oapd.inaf.it}
%\and
%ESO, Karl-Schwarzschild-Str. 2, 85748 Garching bei M\"unchen, Germany\\ 
%email{srandall@eso.org}
\and
Institute of Theoretical Physics and Astronomy, Vilnius University, 
Gostauto 12, Vilnius 01108, Lithuania
%\email{rimvydas.janulis@tfai.vu.lt}
%\and
%Instituut voor Sterrenkunde, KU Leuven, Celestijnenlaan 200D, 3001 Leuven, 
%Belgium ????\\
%\email{roy@ostensen.eu ????}
\and
Department of Physics, University of Warwick, Coventry CV4 7AL, UK
%\email{t.r.marsh@warwick.ac.uk}
\and
Department of Physics and Astronomy, University of Sheffield, 
Sheffield S3 7RH, UK
%\email{vik.dhillon@sheffield.ac.uk}
\and
Instituto de Astrofisica de Canarias, Via Lactea s/n, La Laguna, 
E-38205 Tenerife, Spain
\and
Konkoly Observatory of the Hungarian Academy of Sciences, Konkoly-Thege M. 
u 15-17, 1121, Budapest, Hungary\\
%\email{paparo@konkoly.hu,\,molnar.laszlo@csfk.mta.hu}
%\and
%Wise Observatory, Tel Aviv University, Tel Aviv 69978, Israel\\
%\email{elia@wise.tau.ac.il}
%\and
%Institute of Astronomy, National Central University, Chung-Li, 32054 Taiwan\\
%\email{m929008@astro.ncu.edu.tw ???}
}

\date{Received ... June 2017; accepted ...}

% \abstract{}{}{}{}{} 
% 5 {} token are mandatory
 
\abstract
{V391~Peg (alias HS~2201+2610) is a subdwarf B (sdB) pulsating star that shows
both $p$- and $g$-modes. By studying the arrival times of the $p$-mode maxima 
and minima through the O--C method, in a previous article the presence of a 
planet was inferred with an orbital period of 3.2 yr and a minimum mass of 3.2 
\mjup.
%\citep{silvotti07}.
%In this article 
Here we present an updated O--C analysis using a larger data set 
of 1066 hours of photometric time series ($\sim$2.5$\times$ larger in terms 
of the number of data points), which covers the period between 1999 and 2012 
(compared with 1999-2006 of the previous analysis).
Up to the end of 2008, the new O--C diagram of the main pulsation frequency 
($f_1$) is compatible with (and improves) the previous two-component solution 
representing the long-term variation of the pulsation period (parabolic 
component) and the giant planet (sine wave component).
Since 2009, the O--C trend of $f_1$ changes, and the time derivative 
of the pulsation period (\pdot) passes from positive to negative; 
the reason of this change of regime is not clear and could be related
to nonlinear interactions between different pulsation modes.
With the new data, the O--C diagram of the secondary pulsation frequency 
($f_2$) continues to show two components (parabola and sine wave), like in 
the previous analysis.
Various solutions are proposed to fit the O--C diagrams of $f_1$ and $f_2$,
but in all of them, the sinusoidal components of $f_1$ and $f_2$ differ
or at least agree less well than before.
The nice agreement found previously was a coincidence due to
various small effects that are carefully analysed.
Now, with a larger dataset, the presence of a planet is more uncertain and 
would require confirmation with an independent method.
The new data allow us to improve the measurement of \pdot\ for $f_1$ and $f_2$:
using only the data up to the end of 2008, we obtain 
\pdot$_1$=(1.34$\pm$0.04)$\times$10$^{-12}$ and 
\pdot$_2$=(1.62$\pm$0.22)$\times$10$^{-12}$.
The long-term variation of the two main pulsation periods
(and the change of sign of \pdot$_1$) is visible also in direct 
measurements made over several years.
The absence of peaks near $f_1$ in the Fourier transform and the secondary 
peak close to $f_2$ confirm a previous identification as $l$=0 and $l$=1, 
respectively, and suggest a stellar rotation period of about 40 days.
The new data allow constraining the main $g$-mode pulsation periods of the 
star.}

\keywords{Stars: horizontal-branch -- 
          Stars: oscillations -- 
          Asteroseismology -- 
          Stars: individual: V391~Peg --
          Planets and satellites: detection --
          Planets and satellites: individual: V391~Peg~b}

\titlerunning{The sdB pulsator V391~Peg and its putative giant planet 
revisited}

\authorrunning{Silvotti et al.}

\maketitle

%-------------------------------------------------------------------

\section{Introduction}

V391~Peg was the first case of a post-red giant branch star showing
evidence of the presence of a planet (\citealt{silvotti07} (hereafter SSJ07), 
\citealt{silvotti08}), indicating that giant planets may survive the first 
giant expansion of a star, provided that the orbital distance is large enough.
For V391~Peg~b, a minimum mass of 3.2 \mjup\ was found, with an orbital period 
of 3.2 yr, corresponding to an orbital distance of about 1.7 AU.
The presence of the planet was inferred by measuring the arrival times of the
maxima and minima of the stellar light, given that V391~Peg is a pulsating 
subdwarf B (sdB) star with at least four $p$-mode pulsation periods between 
344 and 354~s \citep{silvotti02, silvotti10}, and a few longer-period $g$-modes 
\citep{lutz09}.
A recent review on hot subdwarfs of spectral type O and B is given by 
\citet{heber16}.

V391~Peg~b is not the first case in which the light travel-time delay is used 
to detect secondary low-mass bodies.
In principle, the timing technique may be used on any star or stellar system 
that has a sufficiently stable clock, which may be given by the oscillations of
the stellar flux in pulsating stars (like in this case), but also radio signals
in pulsars or eclipse timing in eclipsing binaries.
Radio timing was used to detect the first planetary system around the 
pulsar PSR~1257+12 \citep{wolszczan92}. The extremely high precision of 
the radio pulse made it possible to detect PSR~1257+12~b, the Moon-mass 
innermost planet of the system \citep{konacki03}.
Of the planets detected through eclipse timing, the most convincing case is 
given by two circumbinary planets orbiting the pre-cataclysmic binary NN~Ser.
Eight years after the discovery paper (\citealt{qian09}, see also 
\citealt{beuermann10}) and 26 years after the first data, their existence
remains the best explanation for the observed eclipse time variations
\citep{bours16}.
%
%their existence remains compatible with 26 years of data \citep{bours16}.
%
%(\citealt{bours16} and references therein). 
%
Many other detached close binaries show eclipse time variations: 
for some of them, the presence of planets is excluded by dynamic stability 
computations and the periodic O--C trends may be caused by other reasons, 
such as Applegate-like mechanisms (\citealt{applegate92}, \citealt{lanza06}).
However, for some others, the energy required to produce the quasi-periodic 
changes in the quadrupole moment of the secondary star referred to as the
Applegate mechanism, is too high; and the presence of Jovian planets remains
the most plausible explanation \citep{volschow16}.

The idea of using stellar pulsation to measure the reflex motion that is due 
to a companion is not new (e.g., \citealt{barnes75}).
Recently, the high photometric accuracy achievable from space, in particular 
with the \kep\ mission, has led to a renewed interest in this technique
\citep{silvotti11}, and two systematic approaches based on frequency modulation
(FM) and phase modulation (PM, equivalent to the O--C method) were proposed 
(\citealt{shibahashi12}, \citealt{telting12}, \citealt{shibahashi15}; 
\citealt{murphy14, murphy16b}).

However, to detect low-mass (substellar) companions, we need very 
stable pulsators.
When we exclude all the solar-like oscillators, good candidates are the delta 
Scuti stars (\citealt{compton16}; see also recent discovery by 
\citealt{murphy16a}) and compact stars like white dwarfs or sdB stars.
As for white dwarfs, many articles in the literature have addressed this issue 
(e.g., \citealt{kepler91}), but it has become increasingly evident 
that other effects are present that can mimic light travel time effects
in the O--C diagrams of these stars (e.g., \citealt{dalessio15}).
For sdB stars the situation looks more promising, perhaps because these stars 
have a fully radiative envelope, and there is at least one case in which the 
presence of a low-mass stellar companion detected from pulsation timing was
confirmed by radial velocity measurements \citep{barlow11b}.
Another recent case of a pulsation-timing detection of an F5V companion to an 
sdB pulsator is reported by \citet{otani+17}.

After the detection of V391~Peg~b, some other planet or brown dwarf (BD)
candidates orbiting sdB stars were proposed using different detection methods.
From eclipse timing, about one-third of the known detached sdB/sdO+dM 
(dM=M-dwarf) post-common-envelope binaries (PCEB) are suspected to host 
planets/BDs:
%
%HW~Vir \citep{kilkenny03, lee09, beuermann12b},
%HS~0705+6700 (alias V470~Cam, \citealt{beuermann12a, qian13}), 
%HS~2231+2441 (\O stensen et al. 2007),
%NY~Vir \citep{qian12, lee14} and refs therein), 
%
HW~Vir (\citealt{beuermann12} and references therein),
HS~0705+6700 (alias V470~Cam, \citealt{qian13} and references therein),
HS~2231+2441 (\citealt{qian10} and references therein; but see also 
\citealt{lohr14}),
NSVS~14256825 (\citealt{almeida13}; \citealt{hinse14} and references therein), 
NY~Vir (\citealt{lee14} and references therein), 
and 2M~1938+4603 \citep{baran15}.
Interesting explorations on the origin of PCEB (and specifically sdB+MS/BD) 
circumbinary planets can be found in \citet{zorotovic13}, \citet{schleicher14},
\citet{bear14}, and \citet{volschow16}.
Very different planets or planetary remnants with terrestrial radii have been 
proposed from tiny reflection effects detected by the \kep\ spacecraft in 
KIC~05807616 \citep{charpinet11} and KIC~10001893 \citep{silvotti14}.
However, none of these sdB planet/BD candidates has been confirmed with at 
least two independent detection methods.
More robust detections of a few brown dwarfs (BDs) in eclipsing sdB binaries 
(also called HW Vir systems from the sdB+dM protoptype) were obtained 
by combining stellar radial velocities (RVs) with photometric measurements: 
J08205+0008, J1622+4730 and V2008-1753 have companion masses of about 71, 67, 
and 69 \mjup, respectively \citep{geier11,schaffenroth14,schaffenroth15}.
At least two more sdB+BD eclipsing systems were recently found from the OGLE 
survey (Schaffenroth in preparation, private communication).
Finally, two more BD candidates in sdB binaries were found by combining 
radial velocities (RVs) with photometric reflection effects: CPD-64$\degr$6481 
and PHL~457, with minimum masses of 50 and 28 \mjup, respectively 
\citep{schaffenroth14b}.

In this paper we reconsider the case of V391~Peg, for which we have collected
six years of new photometric time-series data, increasing the number of data 
points by a factor of about 2.5. 
The main stellar parameters of V391~Peg are summarized in Table~1.
 We note that the JHK magnitudes are compatible with a single sdB star and do 
not indicate any near-IR excess.
%
%from spectroscopy are: 
%\teff=29,300$\pm$500~K, \logg=5.4$\pm$0.1 in cgs units and 
%$\log$(N(He)/N(H))=-3.0$\pm$0.3 \citep{ostensen01}.
%The UBV Johnson magnitudes from TNG calibrations are 
%
%Assuming an sdB canonical mass of 0.47 \msun, we obtain a radius of 0.23 
%\rsun, a luminosity of 35 \lsun 

In section 2 a short summary of the data acquisition and reduction is given,
including the extraction of the pulsation frequencies.
The analysis of the amplitude spectrum of the $p$-modes at different frequency 
resolutions is presented in section 3.
Section 4 is dedicated to the O--C analysis of the two main $p$-modes.
In section 5 we discuss the presence of the planet in the light of the new
O--C results, including a perspective on future developments.
In section 6 we present an analysis of the $g$-mode amplitude spectrum.
Finally, a summary of our results is given in section 7.

\begin{table}[h]
%\centering
\begin{center}
\caption[]{Stellar parameters.}
{\small
\begin{tabular}{lc}
\hline
U                   & \hspace{1.2mm}$13.35 \pm 0.03^1$\\
B                   & \hspace{1.2mm}$14.35 \pm 0.02^1$\\
V                   & \hspace{1.2mm}$14.57 \pm 0.02^1$\\
J (2MASS)           & $15.17 \pm 0.05$\\
H (2MASS)           & $15.16 \pm 0.10$\\
K (2MASS)           & $15.38 \pm 0.20$\\
\hline
\teff               & $29\,300 \pm 500$ K$^2$\\
\logg               & $5.4 \pm 0.1$ (cgs)$^2$\\
$\log$(N(He)/N(H))  & $-3.0 \pm 0.3^2$\\
M                   & $0.47^3$ \msun\\
R=R(M, $g$)         & $0.23$ \rsun\\
L=L(\teff, R)       & $34$ \lsun\\
M$_{\rm V}$=M$_{\rm V}$(L, BC) & 3.88$^4$\\
d=d(V, M$_{\rm V}$)  & $1\,400$ pc\\
\hline
%\\
%\multicolumn{2}{l}{Notes: $^1$ Our calibration at TNG.}\\
%\multicolumn{2}{l}{\hspace{8.3mm} $^2$ From \citealt{ostensen01}.}\\
%\multicolumn{2}{l}{\hspace{8.3mm} $^3$ SdB canonical mass (assumed).}\\
%\multicolumn{2}{l}{\hspace{8.3mm} $^4$ Absolute V mag assuming a}\\
%\multicolumn{2}{l}{\hspace{10.5mm} bolometric correction BC=--2.95.} 
\end{tabular}
}
\end{center}
{\small Notes: $^1$ Our calibration at TNG.}

{\small \hspace{8.3mm} $^2$ From \citealt{ostensen01}.}

{\small \hspace{8.3mm} $^3$ SdB canonical mass (assumed), see e.g. 
\citealt{heber16}}.

{\small \hspace{8.3mm} $^4$ Absolute V mag assuming a bolometric correction 
BC=--2.95.}
\end{table}

\section{Time-series photometric data: extraction of the pulsation frequencies}

\begin{table} 
\centering
\caption[]{Time-series photometry.}
{\small
\begin{tabular}{lcrc}
\bf Telescope/instrument & \multicolumn{1}{c}{\hspace{-3mm} \bf Observers} & \multicolumn{1}{c}{\hspace{-0mm} \bf \#~runs} 
& \multicolumn{1}{c}{\hspace{-2.5mm} \bf \#~hours}\\
\hline
Previous data (1999-2006)$^1$ &                                            &168 & \multicolumn{1}{c}{\hspace{-0mm} 421.3} \\
\hline
Loiano 1.5m/BFOSC             & \multicolumn{1}{c}{\hspace{-3mm} RS}       & 20 & \multicolumn{1}{c}{\hspace{1.7mm} 75.4} \\
Piszk\'{e}stet\H{o} 1.0m/CCD  & \multicolumn{1}{c}{\hspace{-3mm} MP/LM}    & 14 & \multicolumn{1}{c}{\hspace{1.7mm} 67.5} \\
Moletai 1.6m/CCD              & \multicolumn{1}{c}{\hspace{-3mm} RJ}       & 26 & \multicolumn{1}{c}{\hspace{1.7mm} 79.4} \\
Wise 1.0m/CCD                 & \multicolumn{1}{c}{\hspace{-3mm} EL}       &  6 & \multicolumn{1}{c}{\hspace{1.7mm} 35.7} \\
Lulin 1.0m/CCD                & \multicolumn{1}{c}{\hspace{-3mm} WSH}      &  7 & \multicolumn{1}{c}{\hspace{1.7mm} 24.2} \\
MDM 1.3m/CCD                  & \multicolumn{1}{c}{\hspace{-3mm} MR}        &  7 & \multicolumn{1}{c}{\hspace{1.7mm} 33.4} \\
LOAO 1.0m/CCD                 & \multicolumn{1}{c}{\hspace{-3mm} SLK}       & 47 & \multicolumn{1}{c}{\hspace{-0mm} 134.1} \\
Monet-N 1.2m/CCD              & \multicolumn{1}{c}{\hspace{-3mm} SS/RL}     & 20 & \multicolumn{1}{c}{\hspace{1.7mm} 55.0} \\
Baker 0.6m??/CCD              & \multicolumn{1}{c}{\hspace{-3mm} MR}       &  4 & \multicolumn{1}{c}{\hspace{1.7mm} 11.5} \\
Mercator 1.2m/CCD             & \multicolumn{1}{c}{\hspace{-3mm} R\O+students}      & 24 & \multicolumn{1}{c}{\hspace{1.7mm} 69.8} \\
WHT 4.2m/ULTRACAM             & \multicolumn{1}{c}{\hspace{-3mm} TRM/VSD}  &  7 & \multicolumn{1}{c}{\hspace{1.7mm} 36.7} \\
NOT 2.6m/ALFOSC               & \multicolumn{1}{c}{\hspace{-3mm} R\O}      &  3 & \multicolumn{1}{c}{\hspace{1.7mm} 11.2} \\
TNG 3.6m/DOLORES              & \multicolumn{1}{c}{\hspace{-3mm} RS}       &  8 & \multicolumn{1}{c}{\hspace{1.7mm} 18.7} \\
Calar Alto 2.2m/CAFOS         & \multicolumn{1}{c}{\hspace{-3mm} SS/RL}    & 10 & \multicolumn{1}{c}{\hspace{1.7mm} 25.9} \\
\hline
Tot new data (2007-2012)      &                                            &203 & \multicolumn{1}{c}{\hspace{1.4mm} 644.9$^2$} \\
All data (1999-2012)          &                                            &371 & \multicolumn{1}{c}{\hspace{-1.4mm} 1066.2} \\
\hline
\\
\multicolumn{4}{l}{Notes: $^1$ See SSJ07 Supplementary Information for more details}\\
\multicolumn{4}{l}{\hspace{11.3mm} (a Monet-N run of Nov 2006 was added to that list).}\\
\multicolumn{4}{l}{\hspace{8.3mm} $^2$ This number is smaller than the sum of col.~4 given that}\\
\multicolumn{4}{l}{\hspace{11.3mm} sometimes overlapping data from different telescopes}\\
\multicolumn{4}{l}{\hspace{11.3mm} were averaged using a weighted mean.} 
\end{tabular}
}
\end{table}

\begin{figure}
\includegraphics[width=9.0cm]{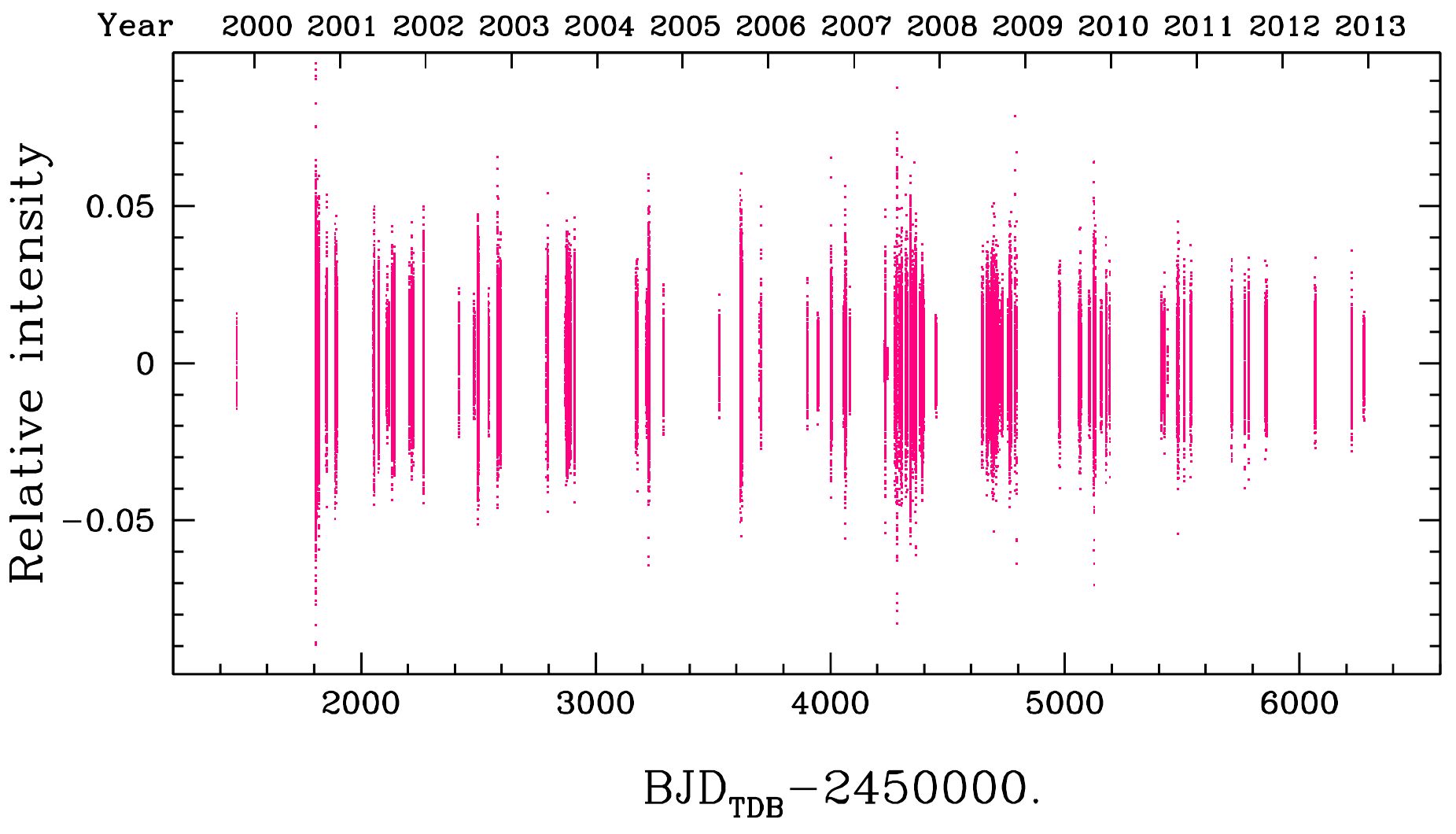}
\caption{Distribution of the 217,232 data points over 13 years.
The overall duty cycle is 0.92\%, and the best coverage is obtained in 2007
with a duty cycle of 5.55\%.
The varying relative intensity is caused by the beating between the main 
frequencies and also depends on the varying quality of the data.}
\label{fig1}
\end{figure}

The new time-series photometric data were obtained using different 
telescopes and instruments (see Table~2) with at least one and often two or 
more comparison stars close to the target in order to remove spurious 
photometric modulations that are due to atmospheric transparency variations.
The distribution of the data during the 13 years of observation is shown in
Fig.~1.
Most of the data were taken using a standard Johnson B filter.
Only at NOT and MERCATOR did we use a Bessell B and a Geneva B filter, 
respectively.
Moreover, a SLOAN g filter was used in the WHT-MDM run of October 
2007\footnote{The WHT data were simultaneously obtained with ULTRACAM in three
photometric bands (u, g, and r) but only the g-band data are 
used in this article, while multi-band data were previously used 
to identify the main pulsation modes of V391~Peg \citep{silvotti10}.}.
The data obtained in October 2007 at the Piszk\'{e}stet\H{o}, Loiano, and 
Lulin Observatories were collected without any filter in order to maximize 
the signal-to-noise ratio (S/N) of that run.
%
%one run with a SLOAN g filter, another one with 
%a Bessell B filter and a third one without any filter.
%
The differences introduced by the different filters in terms of amplitudes
or phases of the pulsation modes were considered and found to be 
negligible because of the much larger volume of standard B measurements.
From nonadiabatic models, these differences (in particular the phase 
differences) are expected to be very small for $l$=0 and $l$=1 modes
(\citealt{randall05}; see in particular their Figs.~13 and 14).
The data were reduced mainly by the observers using standard procedures 
for aperture 
%(and in a few cases PSF) 
differential photometry.
The times of all the data (new and old) were converted into
Barycentric Dynamical Times (BJD$_{\rm TDB}$) following \citet{eastman10}).

From the reduced data we extracted accurate pulsation frequencies using
a classical prewhitening technique: an iterative Fourier transform (FT)
process was applied subtracting the main frequency from the data 
residuals at each iteration, until no frequencies with amplitudes larger than 
four times the FT mean noise level were present.
At the end of this iterative process, the pulsation frequencies, amplitudes,
and phases were optimized through a multi-sinusoidal fit, whose results are 
given in Table~3.
Appropriate statistical weights were set and considered in the sinusoidal fits
of the $p$-modes \citep{silvotti06} in order to take the varying 
quality of the data into account that is due to different telescope apertures, 
instrument efficiencies, and weather conditions.

%If we compare the main four frequencies of Table~3 ($f_1$ to $f_4$) 
%with those reported in Table~2 of \citet{silvotti02}, we see a very good
%agreement for $f_1$ and $f_2$, but some differences for the low-amplitude
%peaks: $f_3$ is now quoted at 2881.12~\muHz\ (it was at 2880.690); 
%for $f_4$ there is a large difference of 11.821~\muHz\ between the new and 
%the old value (2909.995 and 2921.816~\muHz\ respectively), which corresponds 
%to the 1-day alias at 11.574~\muHz. 
%As we will see in the next sections, this is the main reason why the results 
%of the current O--C analysis differ substantially from the previous ones 
%reported in SSJ07.
%Finally, in \citet{silvotti02} a fifth independent low-amplitude frequency at 
%2738.015~\muHz\ was considered, while this frequency is not detected in 
%the new (better quality) data and thus not considered in our current analysis.
%
%
% DECIDERE SE RIMETTERE LA FRASE SUCCESSIVA o SPOSTARE LA SECONDA PARTE 
% AL'INIZIO DELLA SEZ. 4
%
%While for the $p$-modes all the available data were used, for the $g$-modes we
%used only the best quality data corrected for the atmospheric extinction.

\begin{table*} 
\centering
\caption[]{Pulsation frequencies.}
{\small
\begin{tabular}{llllcl}
& & ~~~~~$\overline{\bf F}$~~[\muHz] & ~~~~~~~~$\overline{\bf P}$~~[\rm s] & ~$\overline{\bf A}$~[\rm ppt]$^1$ & \bf ~~~Phase$^2$\\
\hline
\bf \hspace{28mm} \textit{p}-modes$^3$ & $f_1$   & 2860.938272(06) & 349.5356784(07) & 7.56 & 0.7327(06) \\
                              & $f_2$   & 2824.096225(10) & 354.0955832(13) & 4.06 & 0.7492(11) \\
                              & $f_3$   & 2881.123233(62) & 347.0868544(74) & 0.77 & 0.3285(58) \\
                              & $f_4$   & 2909.995332(63) & 343.6431630(75) & 0.65 & 0.2560(58) \\
                              & $f_2^-$ & 2823.932963(57) & 354.1160549(72) & 0.93 & 0.1015(54) \\
\hline
\bf \hspace{28mm} \textit{g}-modes$^4$ & $F_1$   & 201.96312(16)   & 4951.3991(40)   & 1.01 & 0.116(09)  \\ 
                              & $F_2$   & 295.11065(23)   & 3388.5596(26)   & 0.78 & 0.475(12)  \\
                              & $F_3$   & 320.19726(23)   & 3123.0748(22)   & 0.71 & 0.918(13)  \\
\hline
\\
\multicolumn{6}{l}{Notes: $^1$ ppt~=~parts per thousand~=~0.1\%.}\\
\multicolumn{6}{l}{\hspace{8.2mm} $^2$ Normalized phases corresponding to \bjdtdb\ 2451470.476568 (1$^{st}$ datum).}\\
\multicolumn{6}{l}{\hspace{8.2mm} $^3$ For the $p$-modes, frequencies and periods are the mean values in the period 1999-2012, 
corresponding to \bjdtdb\ $\sim$2454090 (or year}\\ 
\multicolumn{6}{l}{\hspace{10.5mm} $\approx$2007.0), which is the weighted mean time.
% from statistical weights. 
We note that in 10 years of observation, the secular variations of the pulsation frequen-}\\
\multicolumn{6}{l}{\hspace{10.5mm} cies and periods are larger than the 1$\sigma$ 
%frequency/period 
errors reported here, obtained from a Monte Carlo simulation assuming constant frequencies.}\\
%\multicolumn{6}{l}{\hspace{10.5mm} ming constant frequencies.}\\ 
%\multicolumn{6}{l}{\hspace{8.2mm} $^4$ For the $g$-modes, frequencies and periods are the mean values over the period 2002-2008.
%The frequency/period errors are under-}\\ 
%\multicolumn{6}{l}{\hspace{10.5mm} estimated and more in general the best multi-sinusoidal fit reported here is much less stable than the
%fit for the $p$-modes because of}\\
%\multicolumn{6}{l}{\hspace{10.5mm} the much higher noise in the Fourier transform at low frequencies (see Fig.~9).}\\
\multicolumn{6}{l}{\hspace{8.2mm} $^4$ Because of the noise in the Fourier transform at low frequencies (Fig.~11), the multi-sinusoidal
fits for the $g$-modes are less stable than}\\ 
\multicolumn{6}{l}{\hspace{10.5mm} those for the $p$-modes, and therefore the 1$\sigma$ frequency/period errors for the $g$-modes 
reported here are underestimated.}\\
\end{tabular}
}
\end{table*}

\section{$P$-modes}

%For what concerns the $p$-modes, frequencies, amplitudes and phases listed in 
%Table~3 were obtained using all the available data.
%
% of the period 1999-2008.
%Data after 2009.0 were not considered because in 2009 we registered a sudden 
%change of $f_1$ (see Fig.~7, 8 and 9), which is discussed in section 3.3.
%
The first problem in analyzing a data set of several 
years is that the pulsation frequencies are no longer constant.
This was already known for V391~Peg, and a quantitative measurement of \pdot\
had been obtained from previous data giving 
\pdot=1.46$\pm$0.07$\times 10^{-12}$ and 
2.05$\pm$0.26$\times 10^{-12}$ for $f_1$ and $f_2$, respectively (SSJ07).
%On the other hand we learned from the results of the Kepler mission
%that most of sdB $g$-mode pulsators show tiny frequency variations with 
%different patterns (see \citealt{zong16} for a detailed study of these 
%variations).
In general, the time variation of a pulsation frequency gradually 
broadens the width of the peak in the Fourier transform and may split it 
into different close peaks if the data set is long enough.
For a linear frequency variation, the time needed to split a pulsation 
frequency into different close peaks is given by
\vspace{-4mm}

%\begin{equation}
%T \approx P~ \left( \frac{c}{\dot{P}} \right) ^{1/2}
%\end{equation}
%
\begin{equation}
T \approx P~ \left( \frac{1.5}{\dot{P}} \right) ^{1/2}
,\end{equation}

\noindent
where P is the pulsation period, and the value 1.5 comes from the actual 
frequency resolution, given by $\sim$1.5/T \citep{loumos77}.
%$\delta$f$\simeq$1.5/T
For V391~Peg we obtain T$\approx$10 years.
%
%where P is the pulsation period and c a constant related to the duty cycle,
%that determines the frequency resolution: $\delta$f=c/T.
%In our case, because of the large data gaps, c$\simeq$1.5 and T$\approx$10 
%years.
However, after a few years, this effect already becomes important and makes the
standard prewhitening technique (which assumes fixed frequencies and 
amplitudes) less efficient in returning precise frequencies.
For this reason, after several tests we decided to split our analysis of the 
amplitude spectrum into three steps with data sets of different length 
and different frequency resolution.

It is useful to recall here that the two main pulsation modes of V391~Peg
were identified as $l$=0 and $l$=1 from high-precision 
multi-color photometry obtained with ULTRACAM at the WHT \citep{silvotti10}.
We show below that this identification is well supported by the current 
analysis.

\subsection{Low-frequency resolution: main pulsation frequencies}

\begin{figure}
\includegraphics[width=9.0cm]{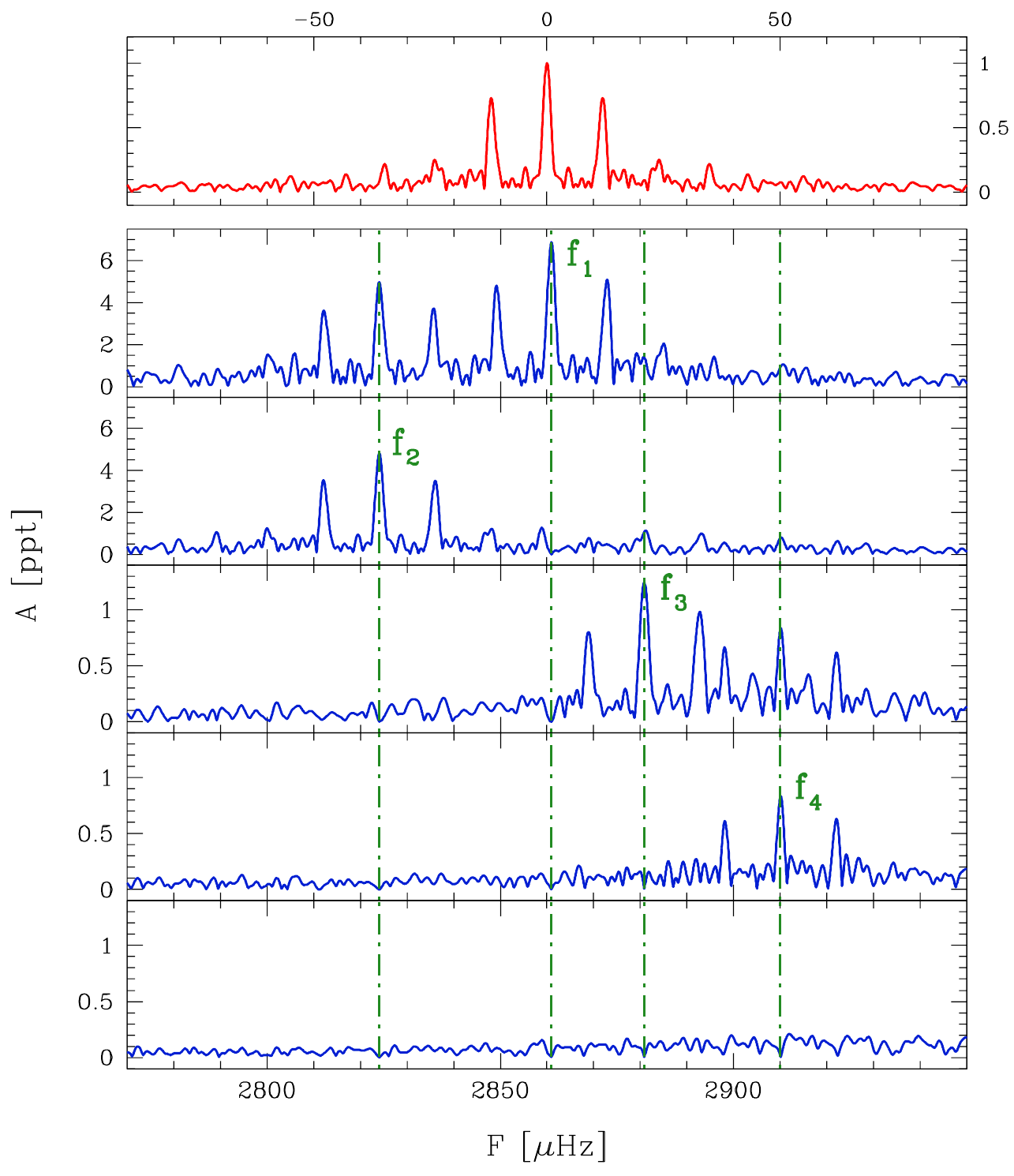}
\caption{$P$-mode amplitude spectrum of our best-quality run of 7.9 days,
with a duty cycle of 35\%, obtained in October 2007 with a SLOAN g filter
using two telescopes at different longitudes: the WHT 4.2m in La Palma,
equipped with ULTRACAM, and the MDM 1.3m at Kitt Peak. 
The upper panel shows the spectral window (red), while the other panels from 
top to bottom show the amplitude spectra of the data and of the residuals 
after one, two, three, and four prewhitening steps.
A plot showing the high quality of the ULTRACAM data is presented
in \citet{silvotti10}.}
\label{fig2}
\end{figure}

\begin{figure}[h!]
\includegraphics[width=9.0cm]{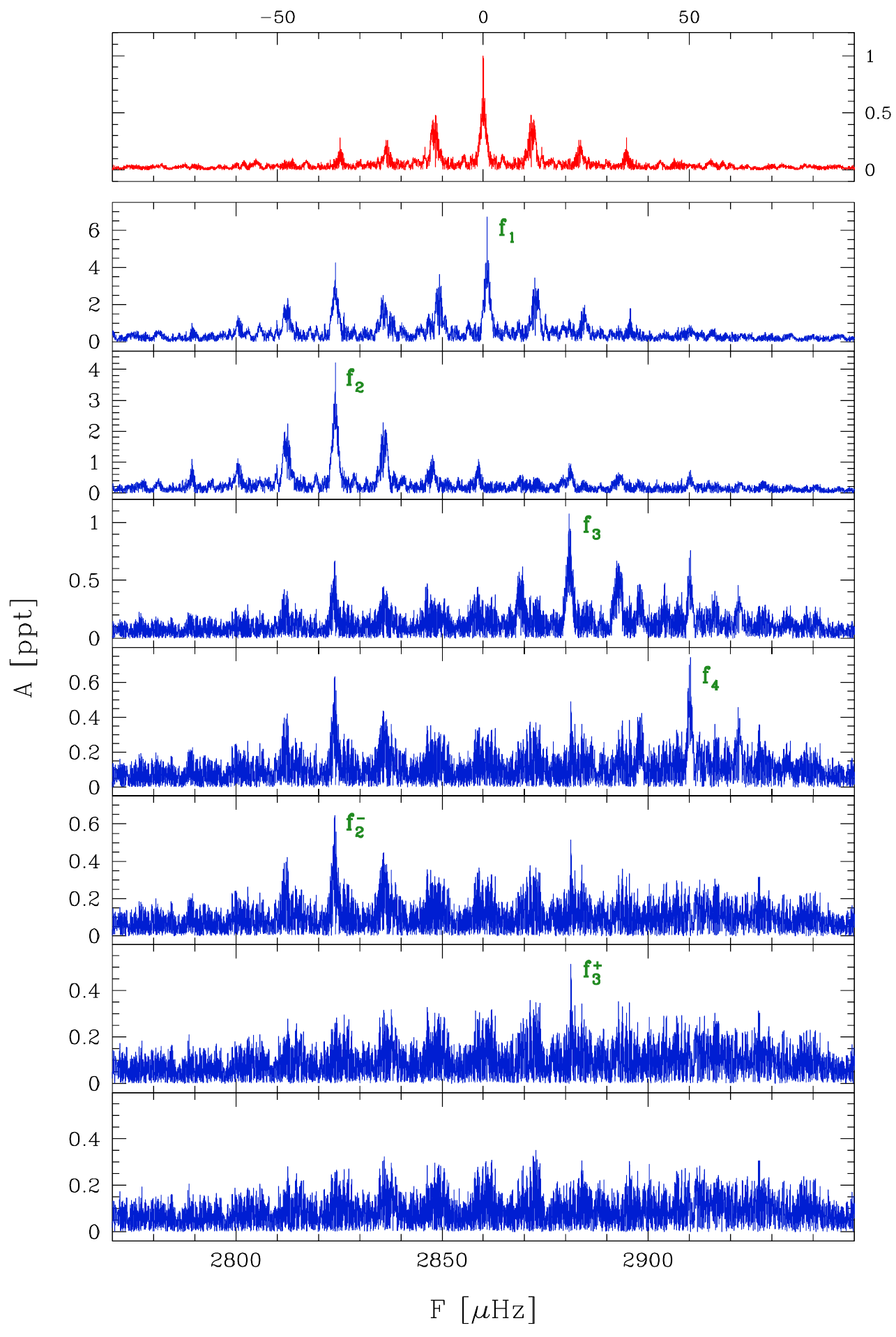}
\caption{Same as Fig.~2, but using all the data of 2007, the year with the best
coverage.
Thanks to the increased frequency resolution, we see that after four 
prewhitening steps, there is still significant power, with secondary peaks
near $f_2$ and $f_3$ that may be due to the rotational splitting of these 
modes.}
\label{fig3}
\end{figure}

\begin{figure*}[t!]
\includegraphics[width=18.14cm]{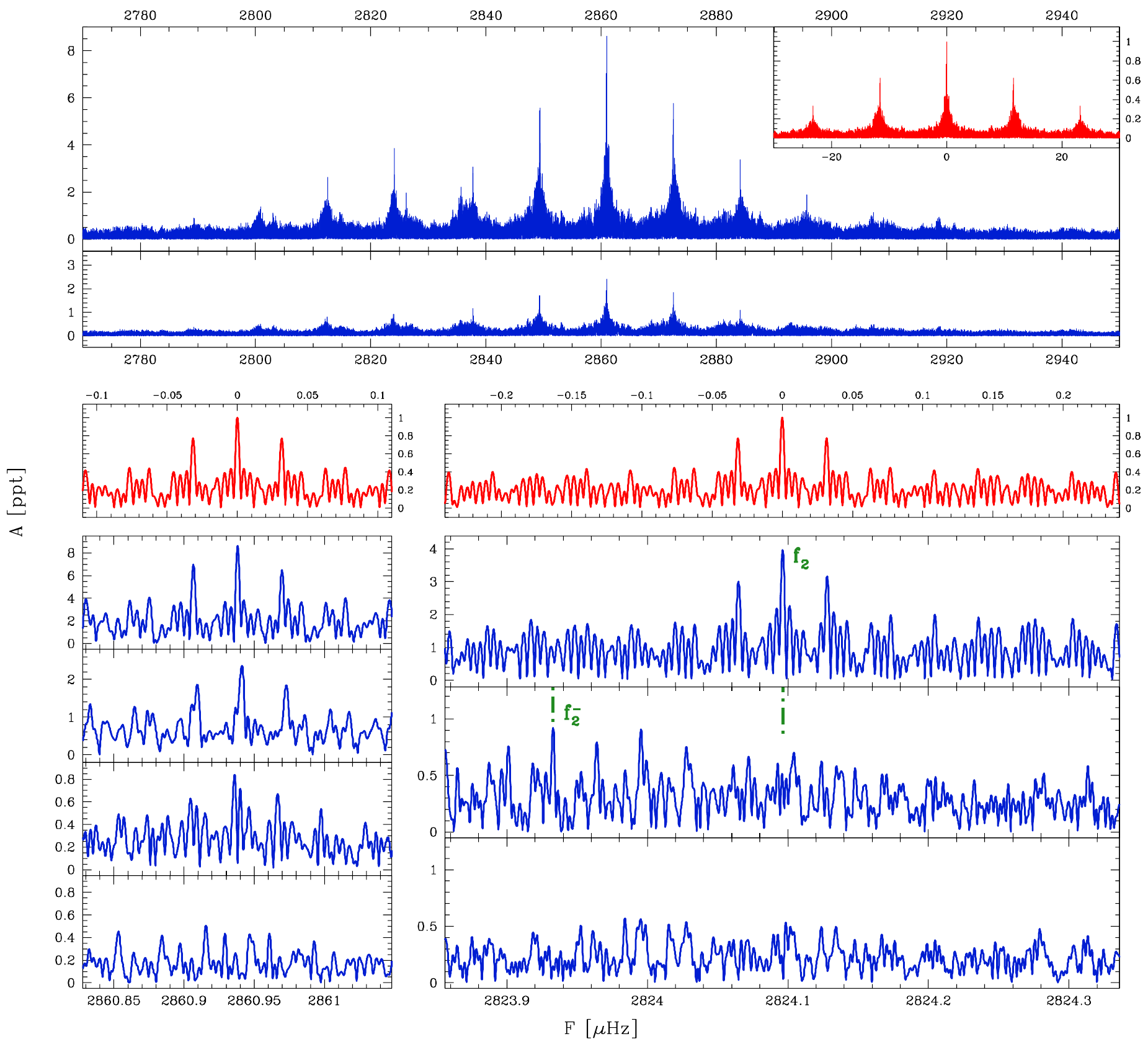}
\caption{Same as Figs.~2 and 3, but using the whole data set (1999-2012). 
Upper panels: amplitude spectrum of the data and of the residuals (on the 
same vertical scale) after subtracting the four main pulsation frequencies 
($f_1$ to $f_4$). 
We note that the residual power is significantly higher than in Fig.~3.
The small box shows the normalized spectral window (red) with the one-day 
aliases at $\pm$11.57~\muHz.
Lower panels (from top to bottom): normalized spectral window (red) with the 
one-year aliases at $\pm$31.7~nHz, and details of the amplitude spectrum of 
data and residuals near $f_1$ (left) and $f_2$ (right). 
The horizontal scale in the left and right panels is the same.
Two vertical dashed lines (green) highlight two components of a possible 
rotational splitting. See text for more details.}
\label{fig4}
\end{figure*}
%
%Left (from top to bottom): normalized spectral window (red in the electronic 
%version) with the 1-year aliases at $\pm$31.7~nHz; amplitude spectrum near 
%$f_1$; residuals after one, two and three prewhitening steps.
%Center: amplitude spectrum and prewhitening for a simulated data set obtained 
%with a single pure sinusoidal wave (no noise) having the same frequency and 
%amplitude of $f_1$ and having also the same long-term frequency and amplitude 
%variations (linear variation of the frequency with 
%\pdot=1.34$\times$10$^{-12}$, and sinusoidal variation of the amplitude like 
%in Fig.~6 upper-right panel).
%Right: same as left panel but for $f_2$; two dashed lines (green in the
%electronic version) highlight a possible rotational splitting. 
%See text for more details.}

As a first step, we consider our best-quality run of October 2007, with a
length of 7.9 days and a duty cycle of 35\%.
At this level of frequency resolution, $\delta$f$\simeq$2.2~\muHz, 
the amplitude spectrum is very clean and shows only four pulsation modes 
without any trace of multiplets of close frequencies (Fig.~2).
%The absence of multiplets, even at higher frequency resolution, was already 
%known \citep{silvotti02}.
%But it's now clear that the main frequencies are only four and not five
%as previously believed \citep{silvotti02}.

\subsection{Medium-frequency resolution: rotational splitting of $f_2$~?}

As a second step, we consider a larger data set of about 220 days, collected
in 2007. This data set is a compromise between best duty cycle, best data 
quality, and relatively long duration in order to detect possible rotational
splitting of the pulsation modes with $l$>0.
At the same time, with 220 days, the effects of the long-term variations 
of the pulsation frequencies are still small, which keeps the amplitude 
spectrum relatively clean (Fig.~3).
%On the same time a duration of 220 days is relatively short to minimize
%the effects of the secular variations of the pulsation frequencies (SSJ07), 
%that make the amplitude spectrum much more complex and more difficult to 
%interpret.
When we removed the four main pulsation frequencies through prewhitening, two
low-amplitude peaks emerged from the noise, close to $f_2$ and
$f_3$, while nothing appeared close to $f_1$, which confirms that this must be 
an $l$=0 mode.
The peak close to $f_3$ ($f_3^+$) 
% differs by +0.527 \muHz\ from $f_3$ but
is only $\sim$3.4$\sigma$ above the noise, which is below our detection 
threshold of 4$\sigma$. Secondary peaks close to $f_3$ are also visible
when we use the whole data set (1999-2012), but with a very low S/N.
The peak close to $f_2$ ($f_2^-$), at about 4.3$\sigma$ above the noise, 
differs by --0.181~\muHz\ from $f_2$ and is also detected in the whole data 
set, but at a lower S/N and smaller separation of
--0.163~\muHz\ (Fig.~4 lower right panel).
Using the latter separation, which is more precise, and assuming that 
$f_2^-$ is part of an $l$=1 triplet split by stellar rotation
in which $f_2$ is the central component, we obtain a stellar rotation 
period of about 40 days.
This value is obtained in the slow rotation approximation 
($\Omega_{\rm ROT}$<<$f$, see \citealt{ledoux51}),

\vspace{-4mm}

\begin{equation}
f_{k,l,m} = f_{k,l,0} + m \, \Omega_{\rm ROT} \, (1 - C_{k,l})
,\end{equation}

%\vspace{-4mm}

\noindent
in which we have used a value of 0.43 for the Coriolis term C$_{k,l}$
according to the adiabatic evolutionary models by \citealt{charpinet02} 
(the model that fits best \teff, \logg\ and P of V391~Peg is model 19 of
sequence 4).
The low amplitude of the secondary peak suggests a low inclination.
This interpretation is consistent with the previous identification of $f_2$
as an $l$=1 mode by \citet{silvotti10}.
A rotation period of $\sim$40 days would be compatible with the 
distribution of rotation periods as recently measured by the \kep\ 
spacecraft in a sample of 18 sdB $g$-mode pulsators (see \citealt{zong17}
and references therein). 
Thirteen of them show periods between 6 and 88 days, with a mean value of 
about 33 days. 
The other five do not show any rotational splitting of the frequencies, 
indicating that they may have very low inclinations and/or extremely long 
rotation periods.

\subsection{High-frequency resolution: frequency and amplitude variations}

\begin{figure}
\includegraphics[width=9.0cm]{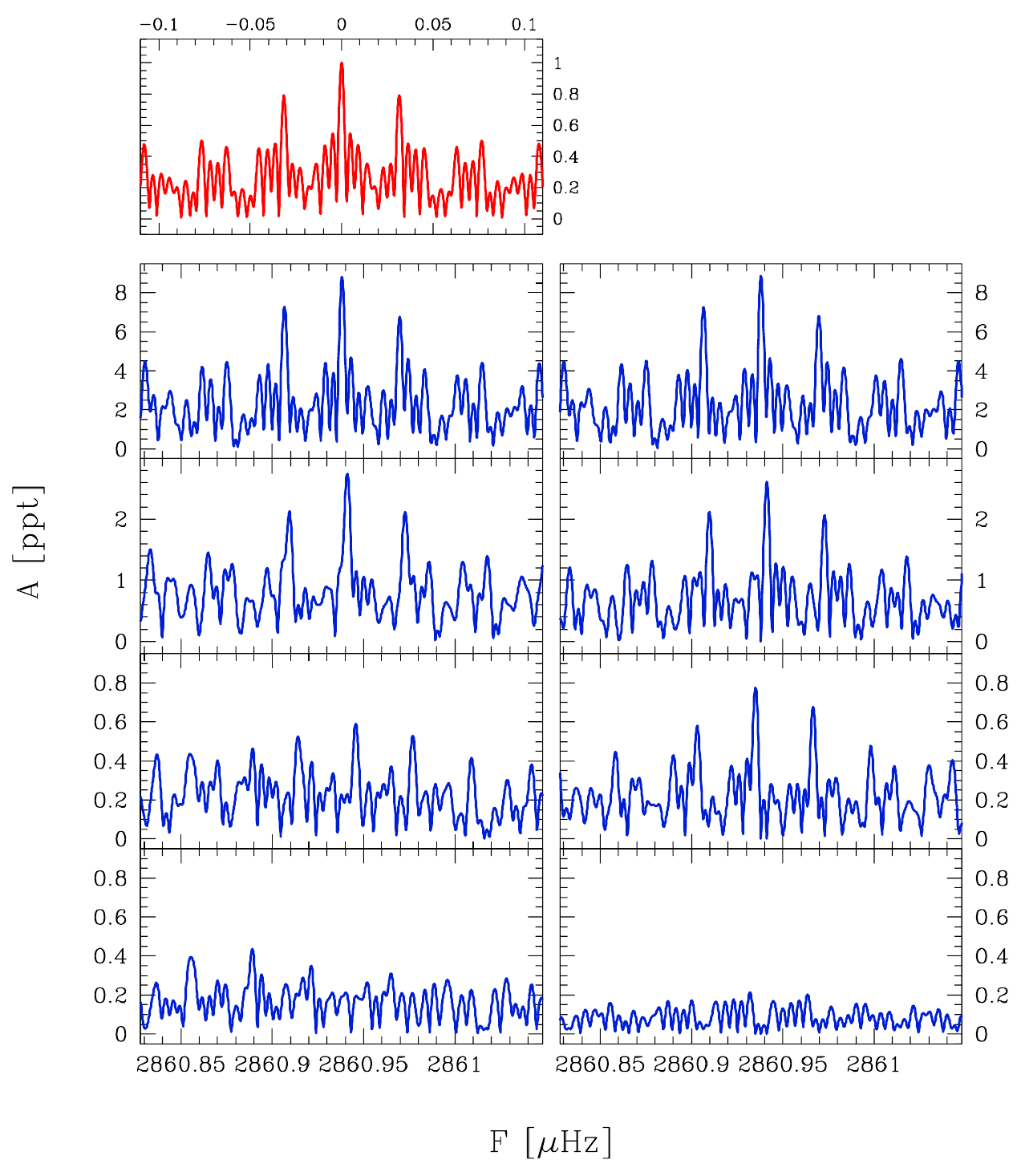}
\caption{Comparison between the amplitude spectrum near $f_1$ of the data 
(left) and the amplitude spectrum near $f_1$ of a simulated data set (right) 
with the same time distribution.
In this test we used only the data up to 2009.0 because in this period 
it is easier to simulate the behavior of $f_1$. 
For the simulated data we used a single pure sine wave (no noise) 
with the same frequency and amplitude of $f_1$ and also with similar 
long-term frequency and amplitude variations (linear variation of the 
period with \pdot=1.34$\times$10$^{-12}$, as derived by the O--C analysis, 
and sinusoidal variation of the amplitude like in Fig.~7 upper right panel, 
green section). Like in the previous figures, the upper left panel is the 
normalized spectral window (red), while the other 
panels are the amplitude spectra of data and residuals after one, two, and 
three prewhitening steps.
This simple test shows that up to the secondary peak on the right side 
of $f_1$, the data are well reproduced by the simulation, both in terms of 
frequency and amplitude. See text for more details.}
\label{fig5}
\end{figure}

\begin{figure*}[ht!]
\includegraphics[width=9.0cm]{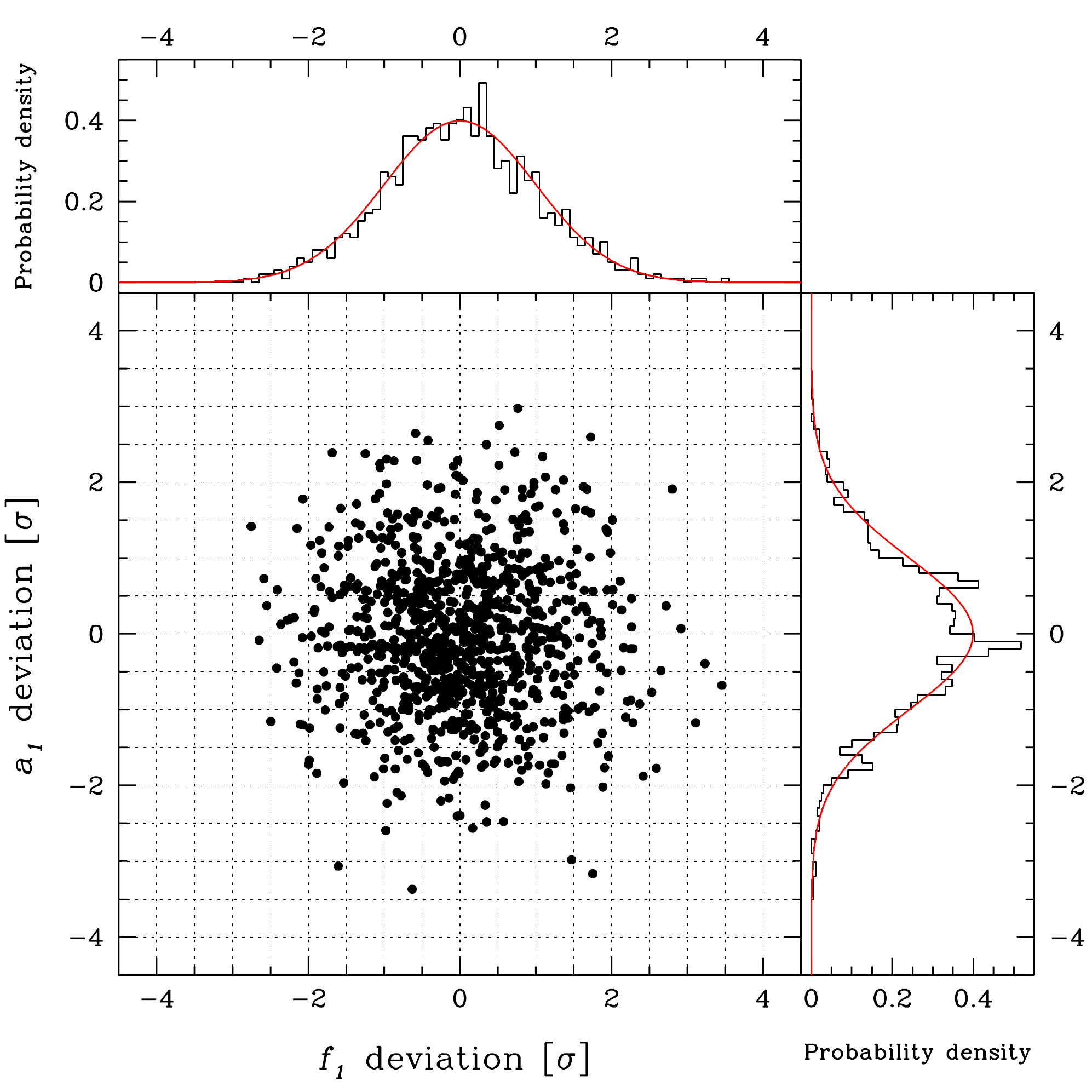}
\includegraphics[width=9.0cm]{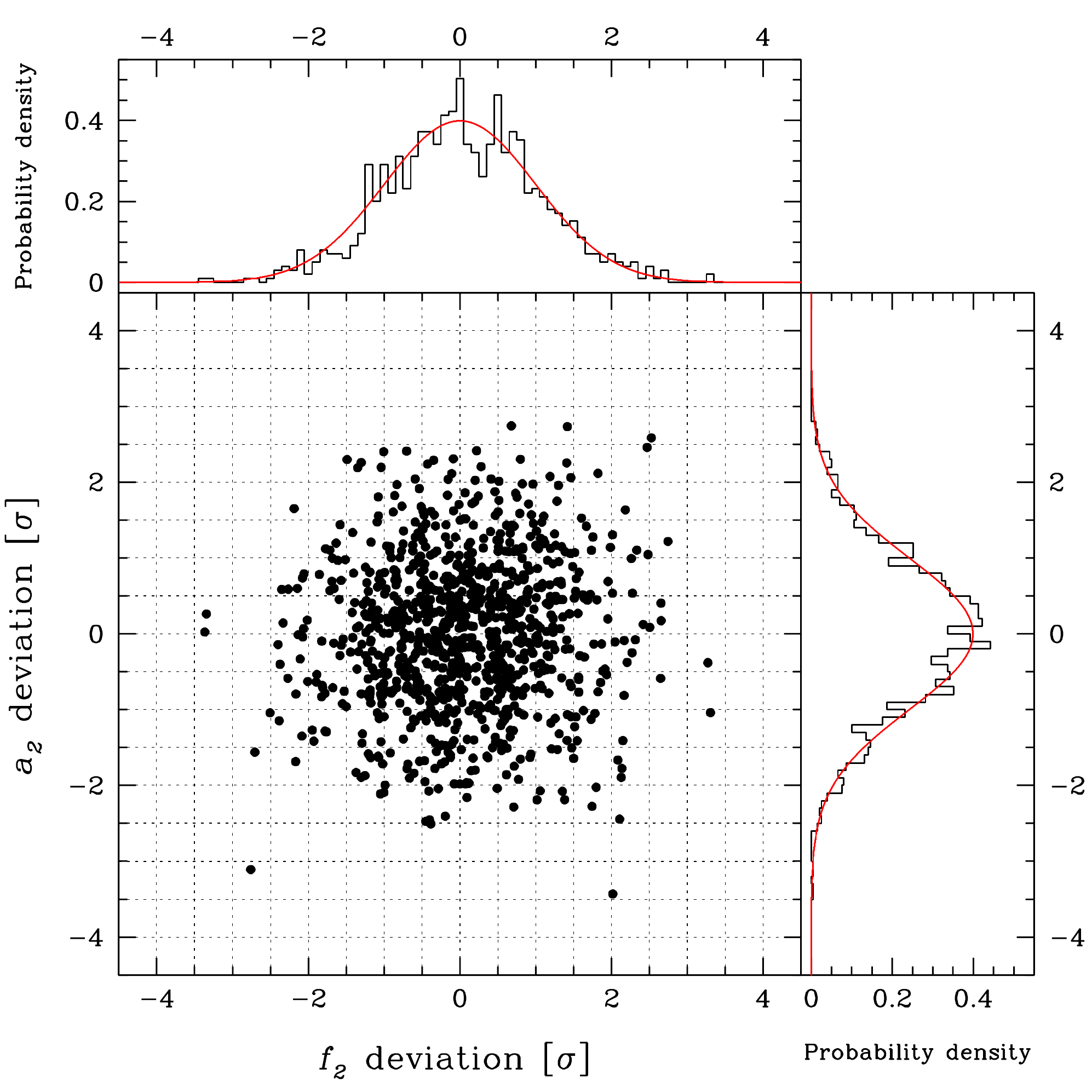}
\caption{Distribution of the frequency and amplitude deviations for the two 
main pulsation modes of V391~Peg.
The deviations, in units of 1$\sigma$ errors, are the differences between 
the values obtained from the original light curve and those obtained from 1000 
artificial light curves created by the MC simulator of Period04 \citep{lenz04}.
The synthetic light curves are built using the five $p$-modes of Table~3 
and adding Gaussian noise at the same level as the original data. 
The 2D distributions are also projected into 1D histograms and compared 
with a normal distribution (red).}
\label{fig6}
\end{figure*}

\begin{figure*}[ht!]
\includegraphics[width=9.0cm]{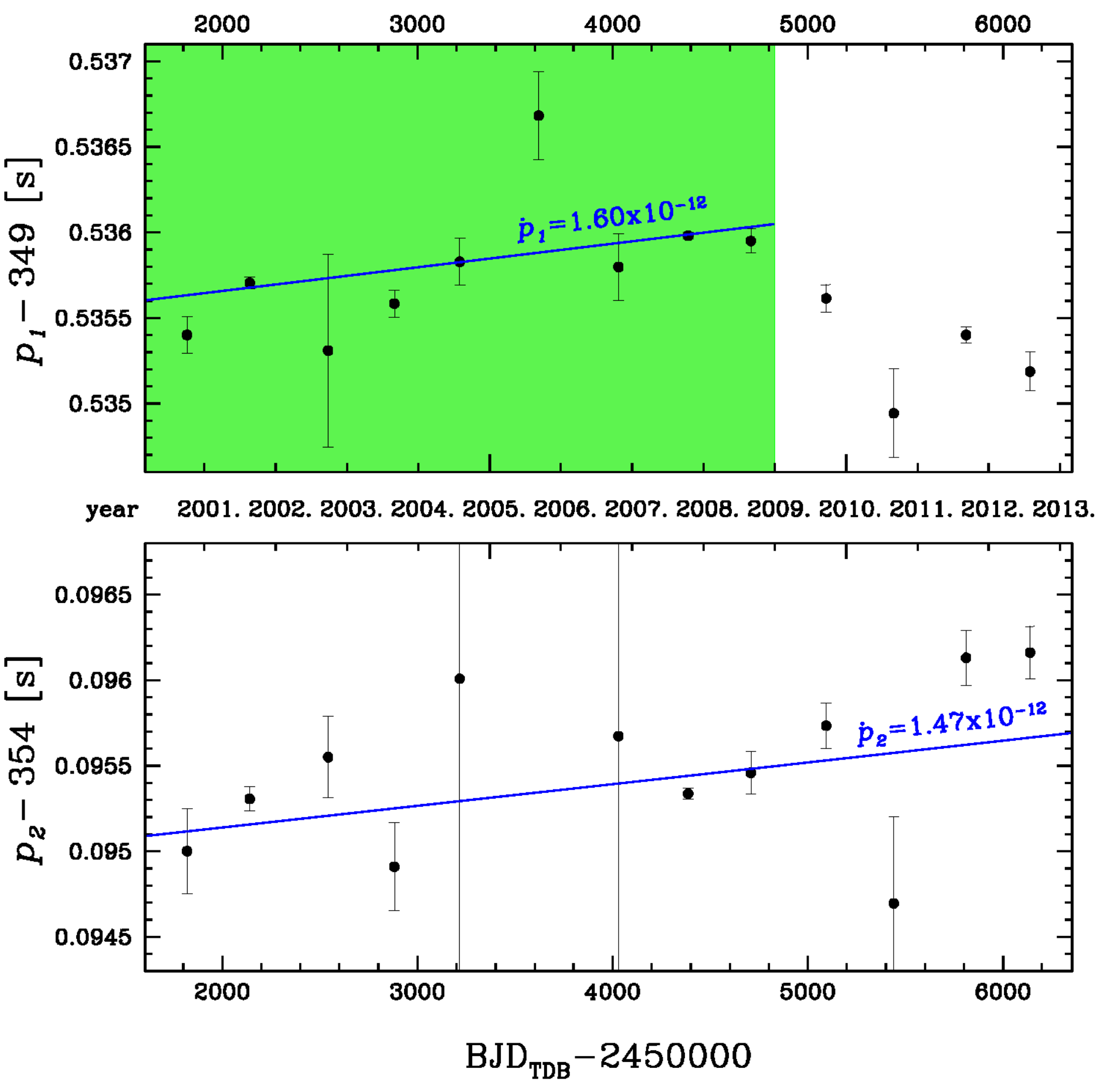}
\includegraphics[width=9.0cm]{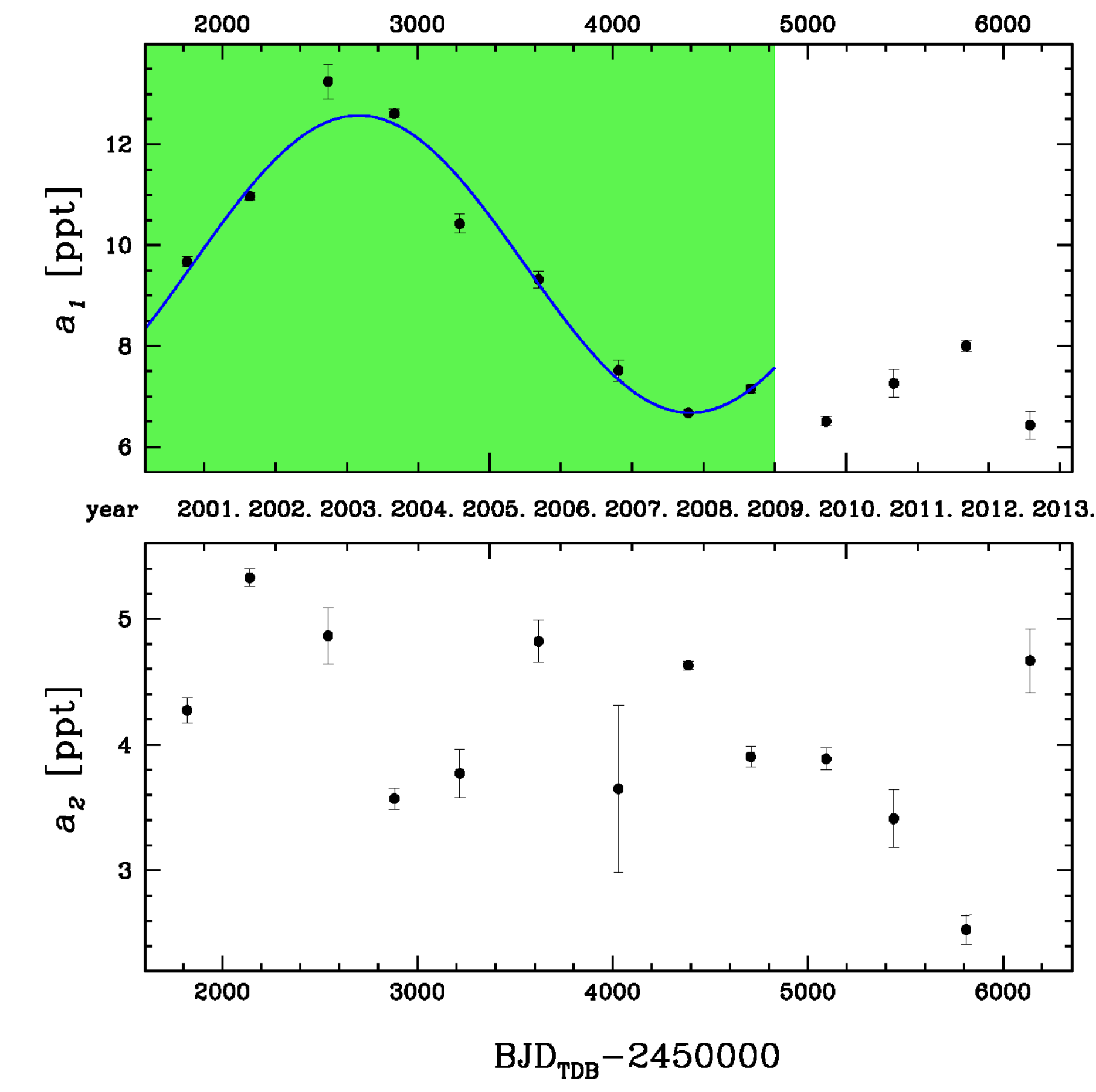}
\caption{Period and amplitude variations of the two main pulsation modes 
of V391~Peg. The variation of $p_1$ is compatible with a linear increase up to
2009.0, when a change of regime appears. The same change is also visible for 
the amplitude: up to 2009.0, $a_1$ shows a fairly regular sinusoidal 
shape with a period of about 3400 days or 9.3 years.
A linear increase of the pulsation period is visible also for $p_2$ when
considering the whole data set, while the irregular variations of $a_2$ can be
at least partially attributed to the beating between $f_2$ and $f_2^-$.
More details are given in the text.
% given that these 
%two close frequencies are not well resolved in most runs.
%
%For $p_2$ and $a_2$, due to the larger error bars, the variations are not 
%significant and they are compatible with a flat behaviour. 
%The lower left plot showing the variations of $p_2$ is split in two panels 
%with different vertical scales for clarity.
}
\label{fig7}
\end{figure*}

When we further increase the length of the data set and consider the
whole light curve in the period 1999-2012, the amplitude spectrum is
much more complex because of the effects of the frequency variations, which
become important (Fig.~4).
When we subtract the main pulsation frequencies from the light curve through
prewhitening, secondary peaks emerge very close to the main pulsation
frequencies.
The reason is that prewhitening subtracts from the data at each step a sine 
wave with constant frequency and amplitude, while on timescales of many years, 
pulsation frequencies and amplitudes are no longer constant.
This effect, which is well visible for $f_1$ (Fig.~4 lower left 
panels), adds noise to the amplitude spectrum of the residuals and may lead 
to incorrect determinations of the low-amplitude frequencies.
In this respect, the average values of $f_3$ and $f_4$ might be slightly 
different from those reported in Table~3, with differences even larger than 
the errors reported there.

In order to decipher the information contained in the peaks close to $f_1$,
we conducted a small experiment with a synthetic light curve.
Since the behavior of $f_1$ is fairly regular and relatively easy to model 
in the period up to 2009.0, while it becomes more irregular later
on (see Figs.~7, 8, and 9), we considered only the period
up to 2009.0.
The synthetic light curve contains a single sine wave without noise with 
the same time distribution as the data, a frequency and amplitude equal to 
$f_1$, and similar frequency and amplitude variations.
In practice, we imposed a linear variation of the period with 
\pdot=1.34$\times 10^{-12}$ (the value found from the O--C analysis described
in section 4) and a sinusoidal variation of the amplitude corresponding to 
the sinusoidal fit shown in Fig.~7 (top right panel).
The amplitude spectrum of this synthetic light curve near $f_1$ is shown 
in Fig.~5 (right panels) and can be compared with the real data in the left 
panels.
Up to the secondary peak on the right side of $f_1$, the agreement between
real and synthetic data is very good both in terms of frequency and amplitude:
we obtain 2860.9418~\muHz\ and 2.74~ppt vs 2860.9414~\muHz\ and 2.61~ppt,
respectively (the main peak being at 2860.9382~\muHz\ with an amplitude of
8.84~ppt).
%
%The amplitude spectrum of the synthetic data shows also a secondary peak on
%the left side of $f_1$ which is not seen in the real data but this peak has 
%a much lower amplitude (<~0.8~ppt) and is much more sensitive to the noise.
%
Thus we verified that a linear time variation of a pulsation period
splits the frequency into three close peaks almost equally spaced in frequency.
If the amplitude is constant, the two secondary peaks have the same amplitude.
If the amplitude is variable as in this case, the two secondary peaks have 
different amplitudes.

Before proceeding with our analysis on frequency and amplitude variations, 
it is important to verify that the uncertainties associated with frequencies 
and amplitudes such as those reported in Table~3 are correctly estimated.
These uncertainties are the 1$\sigma$ errors obtained from a Monte Carlo (MC)
simulation on 1000 synthetic light curves in which random Gaussian noise
(at the same level as the data) was added to the five $p$-modes listed in 
Table~3.
In Fig.~6 the distribution of frequencies and amplitudes obtained from the 
MC simulations is shown for the two main pulsation modes of V391~Peg
($f_1$ and $f_2$).

After we verified that the error bars of our measurements were reliable,
we measured the pulsation periods and amplitudes for $f_1$ and $f_2$
in each observing season (Fig.~7), where observing season means the period
from May to December of the same year in which V391~Peg is observable.
The frequencies and amplitudes shown in Fig.~7 were obtained from
multi-sinusoidal fits considering only four frequencies ($f_1$ to $f_4$), 
while $f_2^-$ was excluded because it is not detected in most of these 
one-season runs. The same exercise was repeated using all five frequencies, 
but the results were less reliable.

\begin{figure*}
\includegraphics[width=18.14cm]{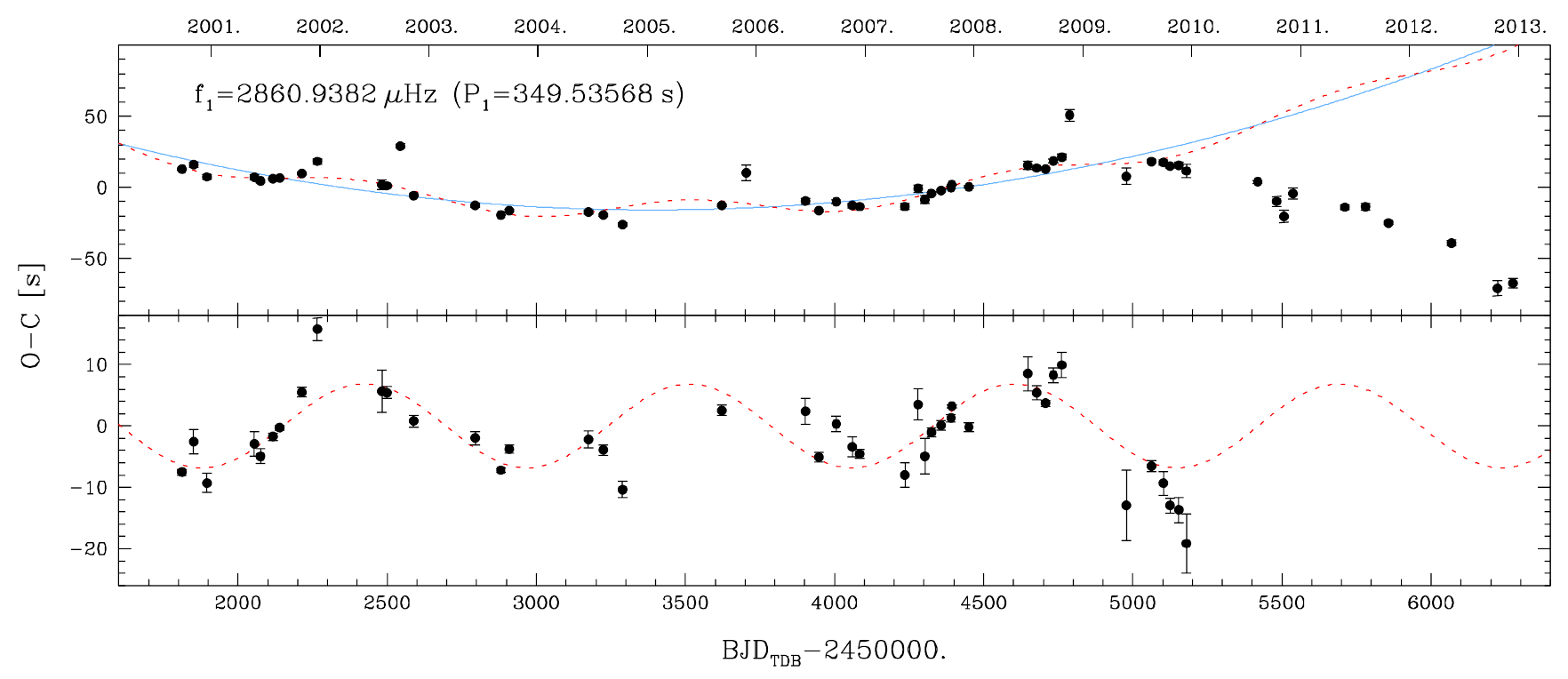}

\includegraphics[width=18.14cm]{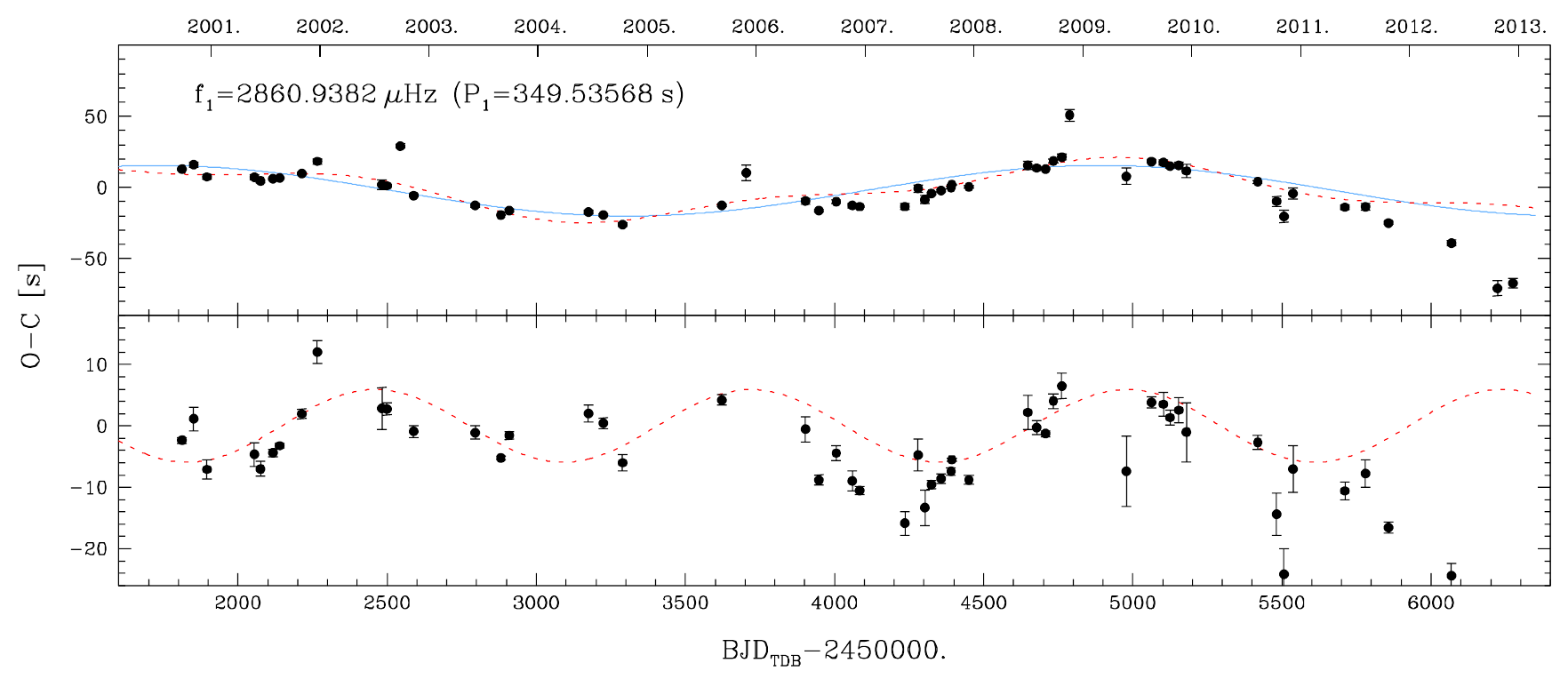}
\caption{O--C diagram of the main pulsation mode of V391~Peg when 
using monthly runs (each point represents the data collected within one month).
Upper panels: fit of the O--C data with a parabola (long-term variation, blue 
continuous line) plus a sine wave (``planetary 
component'', red dashed line) and planetary component alone after subtracting
the long-term component.
This solution gives satisfactory results only up to the end of 2008, and the 
fit was made considering only the data up to 2009.0. 
Lower panels: same as upper panels, but using two sinusoids.
In this case, the fit was made using all the data, but a reasonable fit 
is obtained only up to $\sim$2010, indicating that two components are simply 
not enough to obtain a reasonable fit of all the data.
When we compare the planetary component alone in the period 2000-2009.0,
the fit is better when we use parabola + sine wave ($\chi^2$=762) with respect 
to the double sine wave ($\chi^2$=1267); for comparison, a straight line would 
give $\chi^2$=1376.}
\label{fig8}
\end{figure*}

When we consider only the data up to 2009.0, corresponding to the green part 
of Fig.~7, the variation of 
$p_1$ can be fit with a straight line whose slope corresponds to 
\pdot$_1$=(1.60$\pm$0.20)$\times10^{-12}$.
In the same period, the amplitude $a_1$ shows a fairly regular sinusoidal 
pattern with a period of about 3400 days (9.3 yr) and an amplitude of 29\%.
After 2009.0, the trend of the period and amplitude variations of $p_1$ 
changes and $p_1$ starts to decrease. The reason for this behavior, 
which is also confirmed by the O-C analysis in Figs.~8 and 9, is not known.
Although we normally attribute period and amplitude variations to nonlinear 
interactions between different pulsation modes, in this case, with an $l$=0 
mode, we cannot invoke the resonant mode coupling between the components of 
a multiplet of modes split by the stellar rotation,
nor even the three-mode resonance, which would require that $f_1$ corresponds
to a linear combination of the other two pulsation modes that we do not see.
These two mechanisms were recently invoked as a possible explanation for the
frequency and amplitude variations observed in the sdB $g$- and $p$-mode 
pulsator KIC~10139564 \citep{zong16}.
The lower left panel of Fig.~7 shows that when we use all the available data,
the variation in $p_2$ can be fit with a straight line whose slope corresponds 
to \pdot$_2$=(1.47$\pm$0.41)$\times~10^{-12}$.
In the lower right panel we see quite irregular variations of $a_2$, 
but these apparent variations can be at least partially attributed to the 
interaction (beating) between $f_2$ and $f_2^-$.
When we also consider $f_2^-$ in the fit, the individual measurements of
$a_2$ may vary by several tenths of ppt, indicating that the 1$\sigma$ error 
bars of $a_2$ are underestimated.
At shorter timescales, we did not find any periodicity in the amplitude 
variations of $a_2$ that could confirm the beating effect and thus the 
rotation period of the star around 40 days.
The mean quality of the data is not sufficient for detecting this effect.
Based on our best-quality run of October 2007 at the WHT-MDM, we can only 
exclude short timescale variations (from night to night) for both $a_1$ and 
$a_2$.

We also attempted to fit the data from 1999 to
the end of 2008 with two sine waves corresponding to $f_1$ and $f_2$, leaving 
as free parameters not only the frequencies, amplitudes, and phases, but also 
\pdot$_1$ and \pdot$_2$.
%
%An attempt was done to measure \pdot\ with another method consisting of 
%fitting the data from 1999 to 2008 with two sine waves corresponding to $f_1$ 
%and $f_2$ leaving as free parameters not only frequencies, amplitudes and 
%phases but also \pdot$_1$ and \pdot$_2$.
%
The fit converged only when we fixed \pdot$_2$, but the value that 
we obtained for \pdot$_1$ is about ten times higher than the value obtained 
from the direct measurements.
This method is less reliable than the direct method or the O--C method 
described in the next section because it makes use of constant amplitudes, 
but we know that the amplitudes are not constant, and in particular, $a_1$ 
varies significantly (Fig.~7).

%\vspace{2mm}

While amplitude variations in sdB $p$-mode pulsators have been known for
a long time, with time scales ranging from days to years, the results reported 
in this section show that even the frequencies are less stable than previously 
believed and may suffer significant variations that are not simply due to the 
long-term modifications of the stellar structure.
Amplitude and frequency variations have recently been detected in most of
the sdB pulsators observed by the \kep\ spacecraft, with complex patterns 
that sometimes are stochastic \citep{ostensen14} and sometimes more regular 
and periodic (e.g., \citealt{zong16}).
%
%In a recent detailed study of these variations, Zong et al. (2016) find that 
%these variations are reasonably well interpreted as nonlinear resonant 
%coupling even though the patterns observed are more complex than the 
%theoretical predictions and would require further theoretical work.
%
%For what concerns the $p$-modes, the Kepler contribution has been less 
%important: even in short cadence mode a 58~s sampling is not optimal for 
%pulsation periods that can be as short as $\approx$1 minute.

\section{O--C analysis}

\begin{figure*}[ht]
% \centering
% \resizebox{\hsize}{!}
%\includegraphics[bb=10  10 500  500,width=12cm,clip]{oc1_1999_2006_aa.eps}
\includegraphics[width=9.0cm]{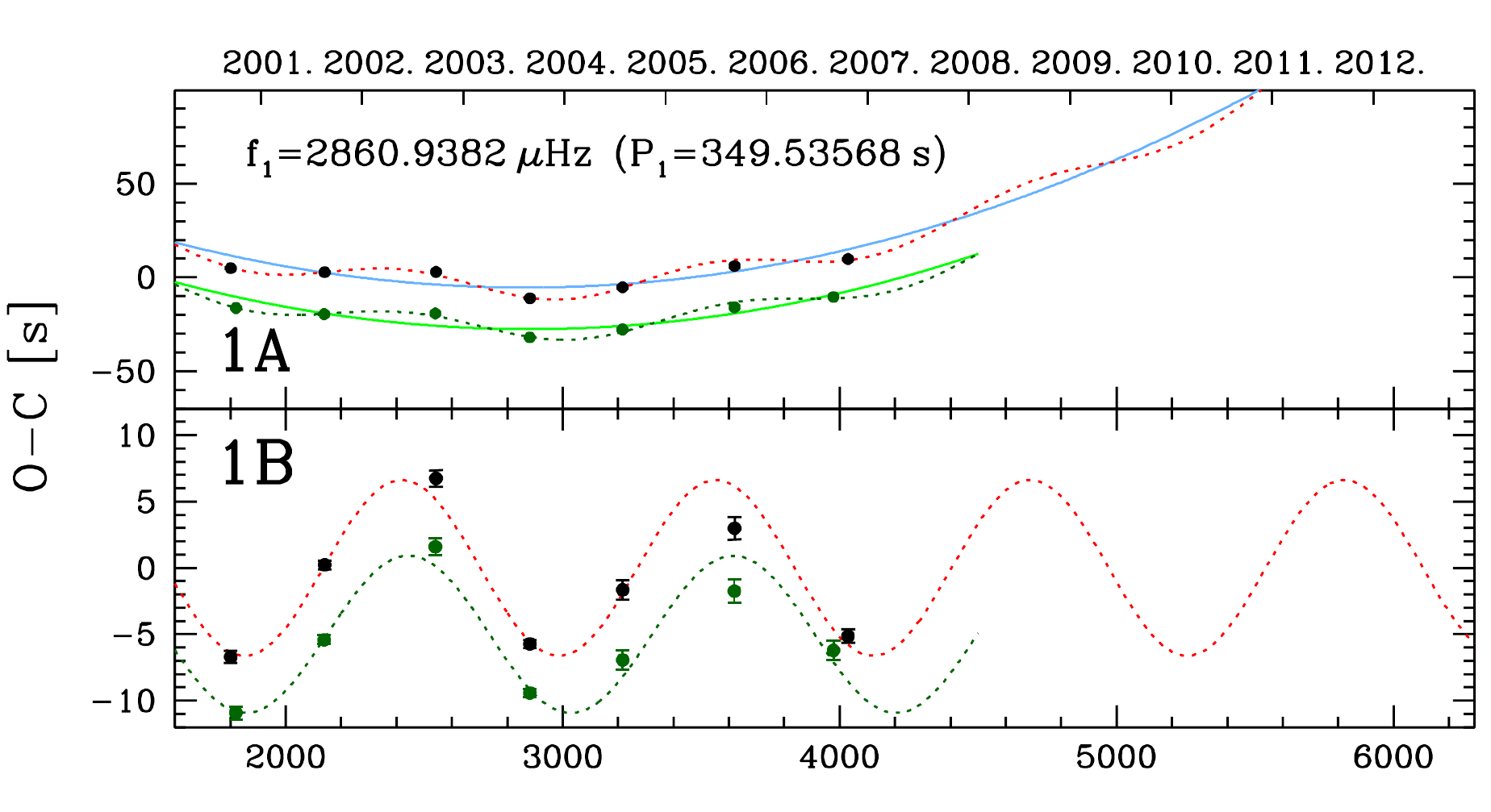}
\includegraphics[width=9.0cm]{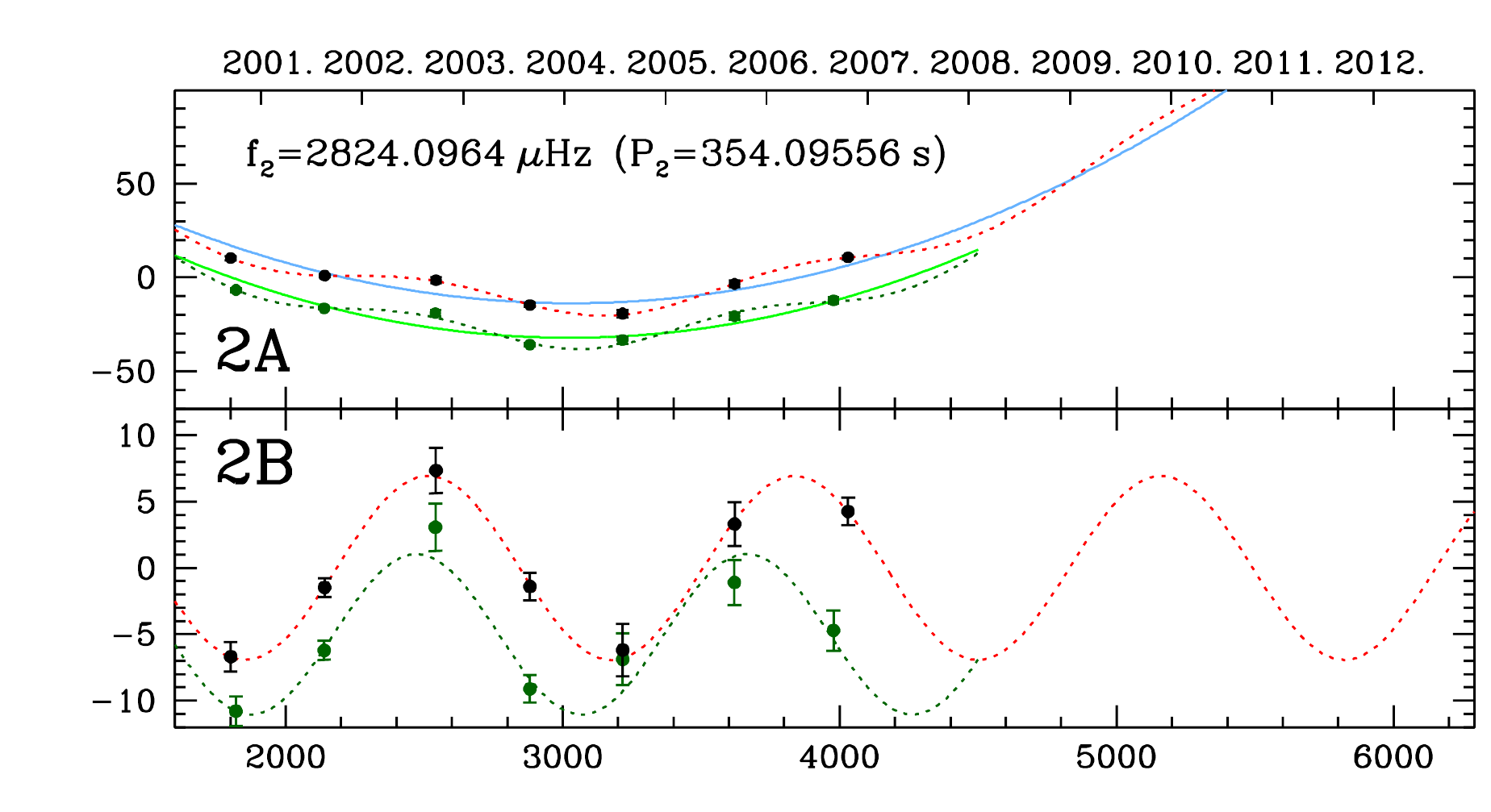}

\includegraphics[width=9.0cm]{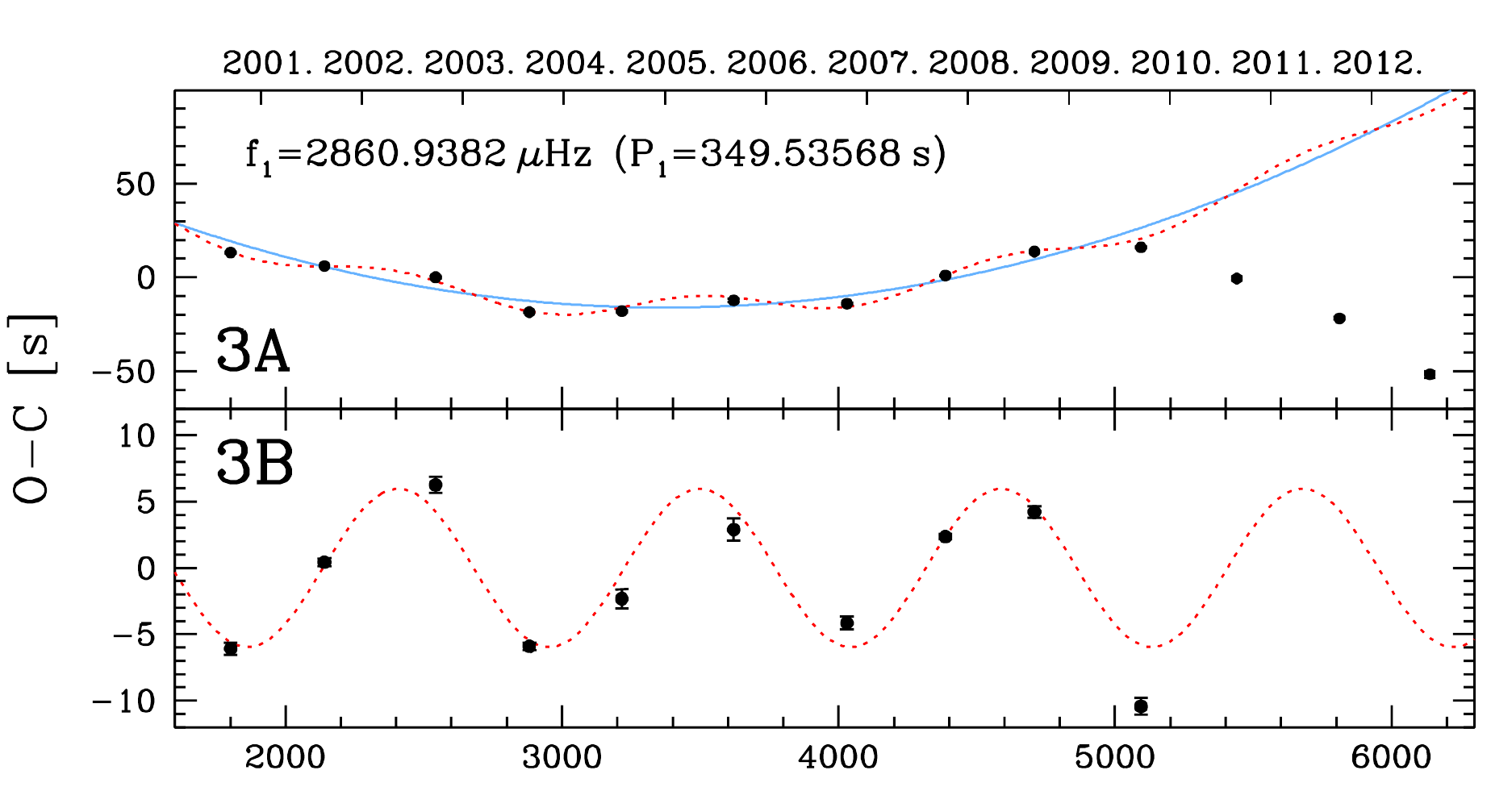}
\includegraphics[width=9.0cm]{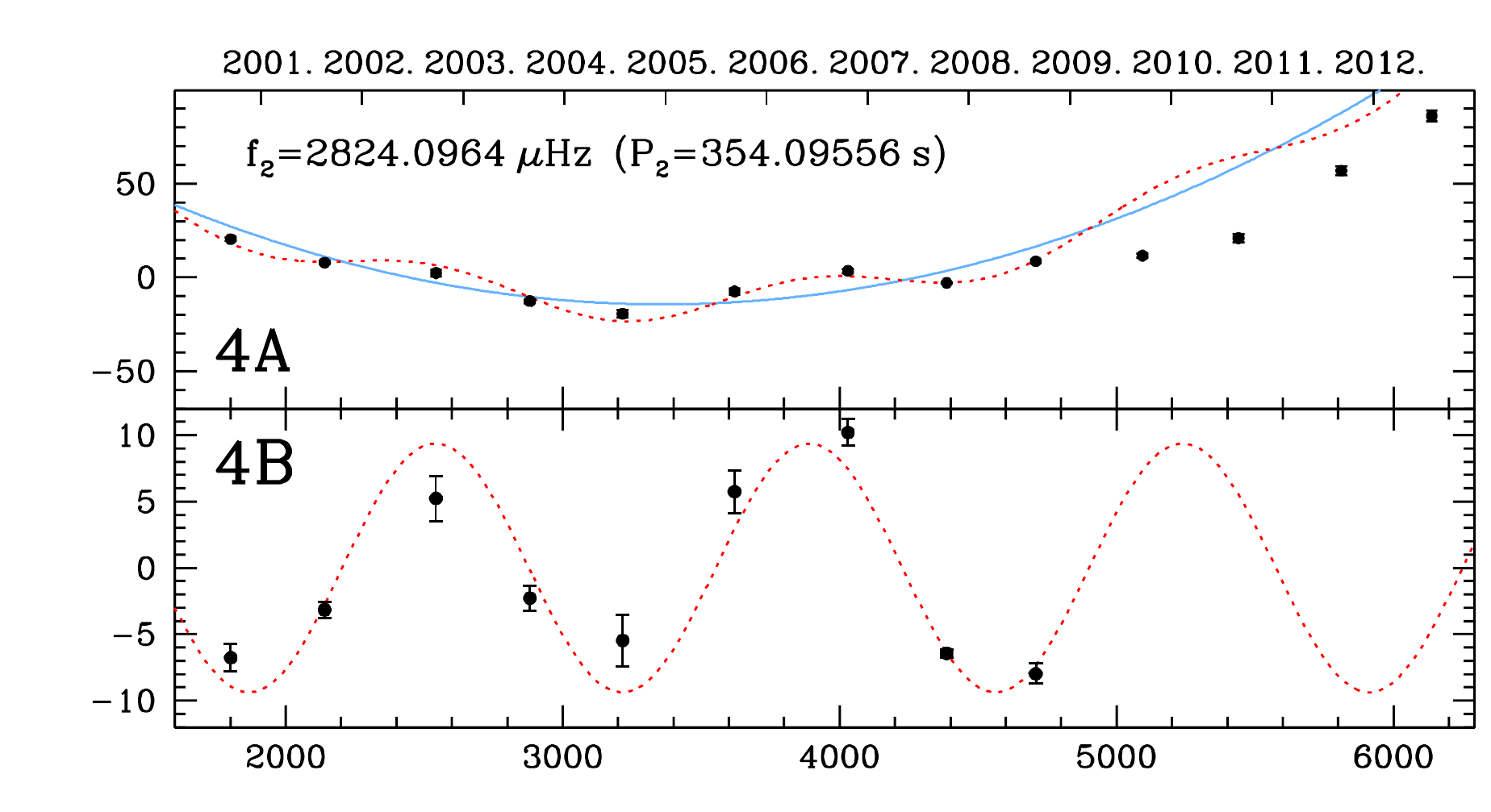}

\includegraphics[width=9.0cm]{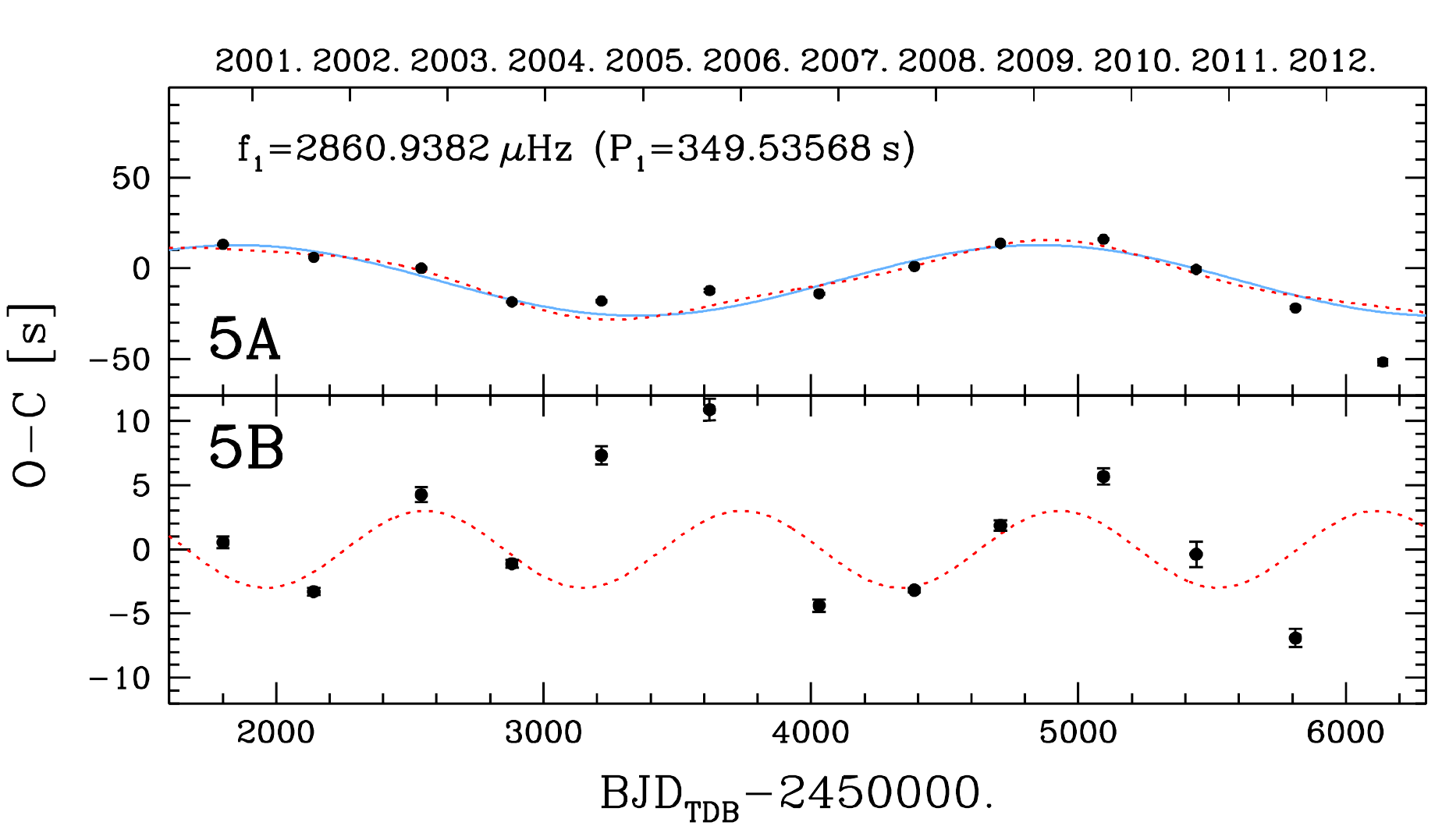}
\includegraphics[width=9.0cm]{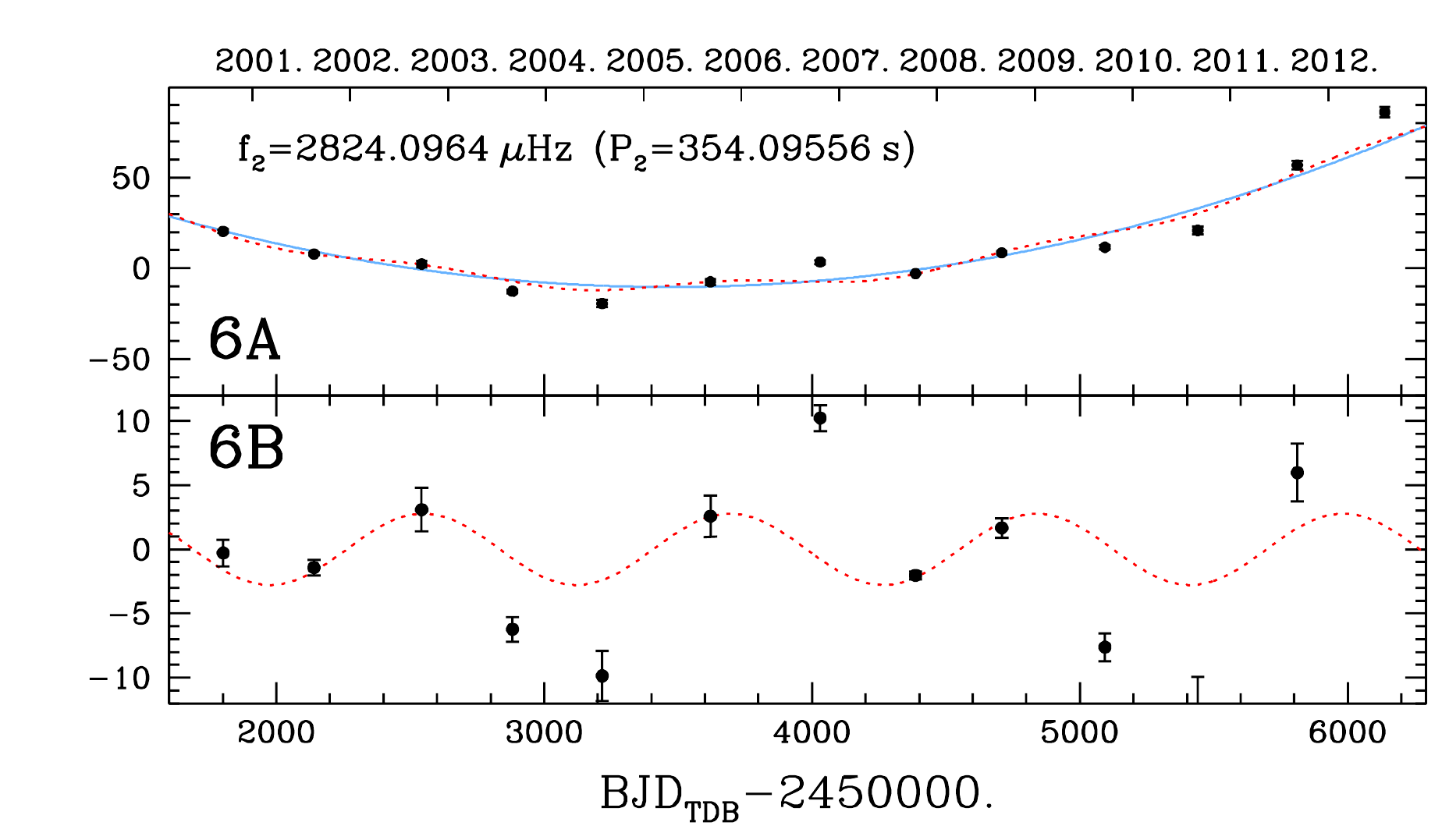}
%\includegraphics[bb=10 510 500 1000,width=12cm,clip]{oc2_1999_2006_aa.eps}
%\vspace{18.0cm}
%\special{psfile=oc1_1999_2006_aa.eps vscale=50 hscale=50 angle=0 voffset=-60
%hoffset=-10}
%\special{psfile=oc2_1999_2006_aa.eps vscale=50 hscale=50 angle=0 voffset=-138
%hoffset=160}
\caption{Same as Fig.~8 for $f_1$ (left) and $f_2$ (right) for 
one-season runs.
Panels 1A, 1B, 2A, and 2B are obtained using only the data up to 2007.0, 
so that we can directly compare the current results (blue and red 
lines) with those obtained by SSJ07 (green lines shifted by -20~s and -5~s 
in panels 1A and 2A and 1B and 2B, respectively).
The small horizontal shifts of the first and last points are due to the 
addition of three observing runs that were not present in SSJ07.
Panels 1B and 2B show that in the current results, the period of 
the sinusoid is slightly shorter for $f_1$ but longer for $f_2$, so 
that at the end the agreement between $f_1$ and $f_2$ is worse with respect 
to SSJ07.
The reasons of these differences are discussed in the text.
When we add the new data, the longer period of the sinusoidal component of 
$f_2$ with respect to $f_1$ is confirmed (panels 3B and 4B), and moreover, we 
note a further difference in amplitude.
Panel 3A confirms the change of regime of $f_1$ near 2009 that was already
visible in Figs.~7 and 8.
This change also tends to worsen the fit of $f_2$ (4A), and for this reason, 
the fits shown in panels 3A to 4B are obtained considering only the data 
up to 2009.0. 
Panels 5A and 5B show an alternative solution obtained using a low-frequency 
sine wave for the long-term component of $f_1$, as in the lower panels of 
Fig.~8.
The fits shown in panels 5A to 6B were obtained using all the available data.
More comments are given in the text.}
\label{fig9}
\end{figure*}

The O--C analysis (\citealt{sterken05}; and subsequent articles in the same 
volume) is a powerful method for detecting tiny variations of the pulsation 
periods on long timescales that cannot be seen or clearly seen from direct 
independent measurements (like in Fig.~7).
The O--C method is more sensitive than the direct method because
instead of directly measuring the period change, it measures the phase 
variations induced by the period change.
When we consider a period that changes linearly in time (a good approximation 
on timescales of a few years, extremely short with respect to the 
evolutionary timescales), the phase variations have the great advantage of 
being proportional to T$^2$, where T is the duration of the observation.

In order to reduce the phase errors, the data for the O--C analysis were 
considered in monthly subsets. 
A four-sinusoid fit was applied to each subset using the best (fixed) 
frequencies from Table~3 ($f_1$ to $f_4$) and leaving amplitudes and phases 
as free parameters.
$f_2^-$ was not used because it is not detected in the monthly subsets.

The difference between these monthly phases and those obtained from the
whole data set are the O-C differences shown in Fig.~8, in which the phase
differences have been converted into time differences.
In Fig.~8 we see the same effect as was already seen in Fig.~7: since 2009, 
the curvature in the O--C diagram of $f_1$ changes.
We do not know the reasons for this change, it might be related to nonlinear 
interactions between different pulsation modes. 
In any case, it is clear from Fig.~8 (upper panels) that a two-component 
fit with a parabola plus a sinusoid (like in SSJ07) can give satisfactory 
results only up to $\sim$2009. 
When considering only the data up to 2009.0, the long-term parabolic variation 
of the main pulsation period corresponds to 
\pdot$_1$=(1.36$\pm$0.06)$\times$10$^{-12}$.
In order to also fit the more recent data, we tried a different approach 
using two sinusoids (lower panels of Fig.~8).
Even in this way, we did not obtain a resonable fit of the whole data set,
and moreover, the quality of the fit up to 2009 is lower, indicating that a 
sinusoidal \pdot\ is not the solution.

As a second step, the O-C analysis was repeated using larger data subsets
covering a whole observing season (that is, from May to December for V391~Peg)
and using the same pulsation frequencies as before. 
Again, $f_2^-$ was not used because it is not detected in almost all runs.
These larger subsets are particularly useful for $f_2$ (the secondary 
pulsation frequency), in order to reduce the phase errors that are very 
large when we use the monthly subsets.
The results are shown in Fig.~9.
In the upper panels (from 1A to 2B), we see the O--C diagram of $f_1$ and 
$f_2$ when using only the data from 1999 to 2007.0, basically the same data
as in SSJ07 (only three short runs were added), but with the new updated 
frequencies.
These plots show that when we use better values for $f_3$ and $f_4$,
the sinusoidal components of $f_1$ and $f_2$ (panels 1B and 2B) differ: 
even if the amplitudes and the initial phases are still in agreement 
(like in SSJ07), the periods are now different.
In the central panels (from 3A to 4B), we see the new fits when we use the
data from 1999 to 2009.0, before the change of sign of \pdot$_1$: 
the sinusoidal components of $f_1$ and $f_2$ (panels 3B and 4B) are similar to 
the previous ones (panels 1B and 2B), except for a larger amplitude for $f_2$, 
which increases the differences between $f_1$ and $f_2$ .
The parabolic components (panels 3A and 4A) correspond to 
\pdot$_1$=(1.34$\pm$0.04)$\times$10$^{-12}$ and
\pdot$_2$=(1.62$\pm$0.22)$\times$10$^{-12}$, in good agreement with the
previous measurements of SSJ07.
These numbers also agree with adiabatic theoretical expectations for the 
secular variation of the pulsation periods \citep{charpinet02}. 
However, the fact that \pdot$_1$ changed sign near 2009 
indicates that in real stars, these processes may be more complicated.
Finally, in the lower panels of Fig.~9 (from 5A to 6B), we show the best 
two-component fits of the whole data set using two sinusoids 
with different periods for $f_1$, and a parabola plus a sinusoid for $f_2$.
Except for the last points, these fits can reproduce the general trend of
the O--C data (panels 5A and 6A), but show a large dispersion, particularly
for $f_1$: the sinusoidal fits in panels 5B and 6B (chi-squared equal to  
894 and 276, respectively) are only slightly better than a simple straight 
line ($\chi^2$=1075 and 322).
At the same time, the two sinusoidal components have similar periods, 
amplitudes, and phases within 4\%, 8\%, and 7\% respectively.

In order to explore this in more detail, we made a weighted average 
of the O--C data in panels 5B and 6B (which means a weighted average
of the O--C data of $f_1$ and $f_2$ after subtracting their 
long-term component).
The result is illustrated in Fig.~10 and shows that when we sum 
the information from $f_1$ and $f_2$, the fit of the sinusoidal component
improves, and at the end, we have 9 points out of 13 that are consistent
with a sine wave with a period of 1127$\pm$45 days (or 3.09$\pm$0.12 years) 
and an amplitude of 3.02$\pm$0.85 light seconds.
Assuming that the sine wave is caused by the planet and 
that the mass of the sdB star is 0.47\msun, these numbers correspond to 
an orbital distance of 1.6 AU and a minimum mass of 1.8 \mjup.

Although not shown in Fig.~9, we also tried to fit the O--C plots of
$f_1$ and $f_2$ with a parabola plus two sinusoids (corresponding to two 
potential planets), but we were unable to find any solution for which the six 
parameters of the two sinusoids were in reassonable agreement between $f_1$ 
and $f_2$.

Several checks were made in order to ensure that the new O--C results 
reported in this section are correct and robust and to understand 
why in SSJ07 periods, amplitudes, and phases of the sinusoidal components of
the O--C diagrams of $f_1$ and $f_2$  agreed so well.
As stated previously, the current O--C results were obtained using four
frequencies ($f_1$ to $f_4$), also including the data taken with filters 
different from Johnson B, and making use of statistical weights.
However, we also tested different combinations without statistical weights,
excluding all the data taken in filters different from Johnson B (see section 
2), and considering only the two main frequencies $f_1$ and $f_2$. 
In all these tests, the results varied little\footnote{When we consider only 
$f_1$ and $f_2$ in the multi-sinusoidal fits instead of four frequencies, 
the results are almost identical to those reported in panels 3A to 4B of
Fig.~9.
When we use only Johnson-B-filter data, the main difference is that the period 
of the sinusoidal component of $f_1$ increases by 7\%. 
When we do not use statistical weights, we obtain the largest difference, with
the amplitude of the sinusoidal component of $f_2$  reduced from
9.4 to 5.4~s, while all other parameters remain about the same.}.
Thus it is not easy to understand the differences between our current results
and those obtained in SSJ07 (even in that analysis, similar tests with 
different combinations were made).
We conclude that the good agreement found in SSJ07 was a coincidence due to
a few small differences between the two analyses: slightly different pulsation
frequencies, two NOT observing runs that were excluded in SSJ07 because they 
were taken with a
Bessell B filter and that are now included (after careful tests of the effects
on phase and amplitude), and one new standard-B-filter Monet-N observing run 
that was not yet available in SSJ07.
Of these factors, the greatest is probably given by the different 
frequencies that were used.
In SSJ07 we used $f_1=2860.9387$, $f_2=2824.0965$, $f_3=2880.6842$, 
$f_4=2921.8463$, and $f_5=2882.0039$~\muHz.
Comparing these values with those in Table~3, we see very small differences
for $f_1$ and $f_2$, compatible with real period variations; the new value 
of $f_3$ is higher by 0.4390~\muHz; $f_5$ is not confirmed and used
not at all in the new analysis, but its influence must be small because 
of the very low amplitude. 
Finally and mostly important, the updated value of $f_4$
is lower by 11.8510~\muHz\ with respect to the old value, which means that 
in SSJ07, because of the poorer spectral window, we used an incorrect value
corresponding to the one-day alias on the right side of the correct peak.
This is probably the mean reason of the different results.
An incorrect value of $f_4$ can modify the multi-sinusoidal fits and thus 
slightly modify the phases of $f_1$ and $f_2$ as well.

\begin{figure}
\includegraphics[width=9.0cm]{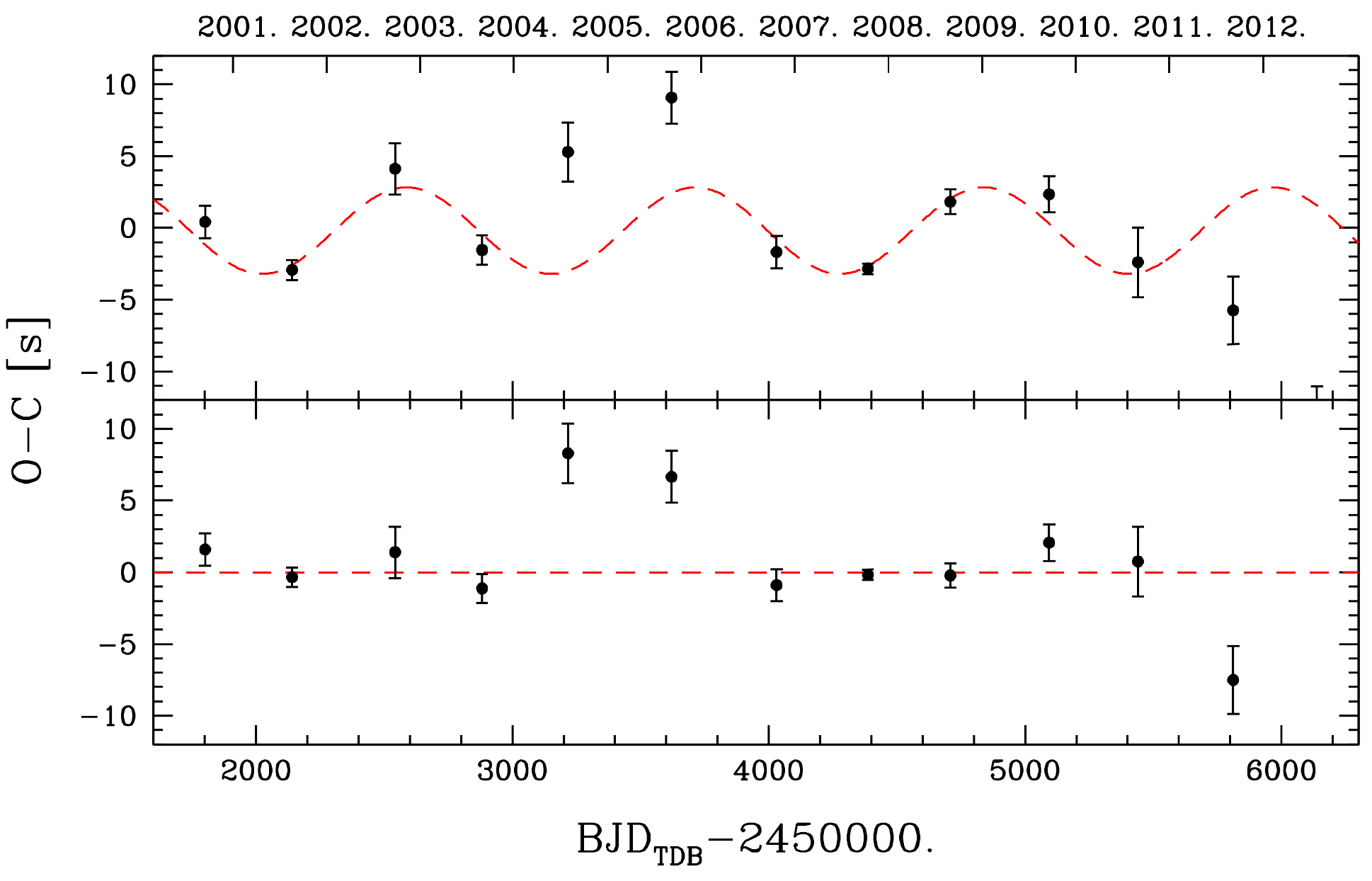}
\caption{O--C diagram obtained by combining the information from $f_1$ and 
$f_2$. In practice, we have computed the weighted average of the points in 
panels 5B and 6B of Fig.~9 and recomputed the best fit with a sine wave.
Compared with these panels, the fit is significantly improved and the residuals
of 9 points out of 13 (including all those with smaller error bars) are close 
to zero.}
\label{fig10}
\end{figure}

\section{V391~Peg~b: real planet or false detection\,?} 

Whether V391~Peg~b is a real planet or a false detection is an open question.
The O--C diagrams of $f_1$ and $f_2$ provide arguments in favour 
and against the presence of V391~Peg~b.

\noindent
1) $f_1$: considering the period up to 2009.0, the O--C diagram of $f_1$ still 
has a sinusoidal component that can be explained by the presence 
of a giant planet with a minimum mass of 3.5 \mjup, orbiting V391~Peg in 3.1 
years at a distance of 1.7 AU.
However, the behavior of $f_1$ after 2009.0 shows that this is more complex,
and we see from Fig.~8 and 9 that a simple two-component fit of the O--C data 
is not enough to interpret the whole data set up to 2012.
Using two sinusoids with different periods allows us to fit the O--C data 
up to 2010 or 2011, but the quality of the fit is much poorer. When we use
two sinusoids, the period of the sine wave corresponding to the planet 
(Fig.~9/5B) is longer than the period obtained with a parabola plus a sine wave
(Fig.~9/3B).

\noindent
2) $f_2$: up to 2009.0, the O--C diagram of $f_2$ also shows a sinusoidal 
component, but now, unlike SSJ07, the period and the amplitude differ from 
$f_1$ by $\sim$20\% and $\sim$36\%, respectively.
The new data support the previous identification of $f_2$ as an $l$=1 mode, 
and this implies that frequency splitting due to stellar rotation must 
be at work.
Regardless of whether our detection of $f_2^-$ is real, 
these modes split by stellar rotation must be there, close to $f_2$, and 
this is a source of noise for the O--C computations of $f_2$.
This argument makes the O--C results from $f_1$ (which is an $l$=0 mode)
more reliable, and this is one of the reasons why the presence of the planet 
cannot be excluded.
At the same time, this argument can partially explain the discrepancies 
between the O--C diagrams of $f_1$ and $f_2$.

\noindent
3) $f_1$+$f_2$: when we try to fit the whole set of O--C data using a
sine wave plus a longer-period sinusoid for $f_1$ and a parabola for $f_2$
(panels 5 and 6 of Fig.~9), we see that the sine wave corresponding to the 
planet is very similar for $f_1$ and $f_2$ in terms of period, amplitude, and 
phase (panels 5B and 6B of Fig.~9).
Although these fits are of poor quality, it is possible to obtain a substantial
improvement when we use both pulsation frequencies together (Fig.~10).
If we interpret this effect with the presence of the planet, we obtain
a minimum mass of 1.8 \mjup, while the orbital period and distance, 
3.1 years and 1.65 AU, do not change much with respect to the 
values obtained previously.

In conclusion, while in SSJ07 the presence of a planet orbiting V391~Peg
was robustly and independently suggested by the two main pulsation modes
of the star, these two modes now give contradictory indications.
A sinusoidal component is still visible in the O--C diagrams of both $f_1$ 
and $f_2$, but the parameters of the two sinusoids are different in general.
The presence of a planet orbiting V391~Peg is clearly much less robust than 
before, although it cannot be entirely excluded.

The peculiar behavior of $f_1$ with a quite sudden change of sign of its
time derivative after 2008 suggests that pulsation timing is a delicate 
method, with aspects that are still unclear and are likely related to nonlinear
pulsation effects.
As a consequence, the reliability of the O--C method to find low-mass 
companions should be questioned, without forgetting, however, that for sdB 
stars we have at least two cases in which the presence of a stellar 
companion was detected through pulsation timing \citep{barlow11a,otani+17},
and in one case, for CS~1246, this detection was confirmed by 
radial velocity (RV) measurements \citep{barlow11b}.
With respect to V391~Peg, the O--C detection was easier in both cases
because of the much higher companion mass, and for CS~1246,
also because of the much shorter orbital period of $\sim$14 days, which meant 
no problems with the long-term variation of the pulsation period.
Unlike CS~1246, which exhibits a single large-amplitude radial mode, and
EC~20117-4014, which shows three low-amplitude pulsation modes with frequency
separations of $\sim$250 and $\sim$680~\muHz\ \citep{otani+17}, with V391~Peg
we have the additional difficulty that all four pulsation modes are 
concentrated within 86~\muHz, which makes it more difficult to measure 
the phases accurately.

In order to confirm or definitively reject the presence of V391~Peg~b, 
an independent confirmation with another method is needed.
Given that Gaia astrometry is not accurate enough at a distance of about 
1400~pc, spectroscopic RVs seem the most natural way to proceed.
However, the RV ``noise'' produced by the pulsations is a serious concern
and can easily reach several hundred m/s, while the expected planetary signal 
is no more than 100 m/s.
Given the very different time scales, it is in principle possible
to remove or reduce the noise due to the 
pulsations, provided that we know the 
Fourier spectrum and the main pulsation modes in detail.
This is true for the high-frequency part of the spectrum (the $p$-modes),
which is relatively simple, with only two dominant modes that
have similar periods.
The noise due to the $p$-modes can be reduced by choosing an exposure time 
close to an integer multiple of $\sim$350~s.
For the $g$-modes, the situation is more complicated as the low-frequency part
of the Fourier spectrum is not well known (see next section).
The noise can be reduced by averaging the results obtained from different 
spectra taken in the same epoch at different pulsation phases.
A great help for a precise determination of the $g$-modes may come 
from TESS (Transiting Exoplanet Survey Satellite, \citealt{ricker16}), 
which can observe V391~Peg continuosly for 54 days in some years 
from now, with a sampling time of 20 or 120~s.
%At that time two high-resolution and high-stability spectrographs will be 
%available on 8m-class telescopes, PEPSI@LBT and ESPRESSO@VLT.

\section{$G$-modes}

$G$-modes were detected in V391~Peg by \citet{lutz09}.
Our new larger data set has been used to confirm this detection.
Given that the $g$-modes are particularly disturbed by the atmospheric 
variations that act at similar frequencies, we selected a subset of 
high-quality data with a length of each single run of at least a few hours.
This subset, which has a total duration of 192.8 hours spread over 5.8 years 
(between 2002 and 2008), was corrected for differential atmospheric 
extinction (the comparison stars are always much redder than the sdB) and 
analyzed. The amplitude spectrum in Fig.~11 shows two regions with an excess 
of power near 180 and 310~\muHz\ and three peaks that emerge from
the noise at more than 5$\sigma$. 
The corresponding frequencies, amplitudes, and phases are listed in Table~3.
The noise threshold, which was 4$\sigma$ for the $p$-modes, was increased 
to 5$\sigma$ because the spectrum is much more noisy in this region.
After these three peaks were subtracted from the data, the lower 
panel of Fig.~11 shows that some residual power is still there, suggesting that
further low-amplitude frequencies are likely present below the noise 
threshold.
As anticipated in the previous section, in two years from now, TESS will be
able to shed light on this part of the Fourier spectrum and likely measure
the rotation period of the star, confirming or refuting the tentative rotation 
period of $\sim$40 days suggested by the $p$-mode analysis in section 3.2.

\begin{figure}[t]
\includegraphics[width=9.0cm]{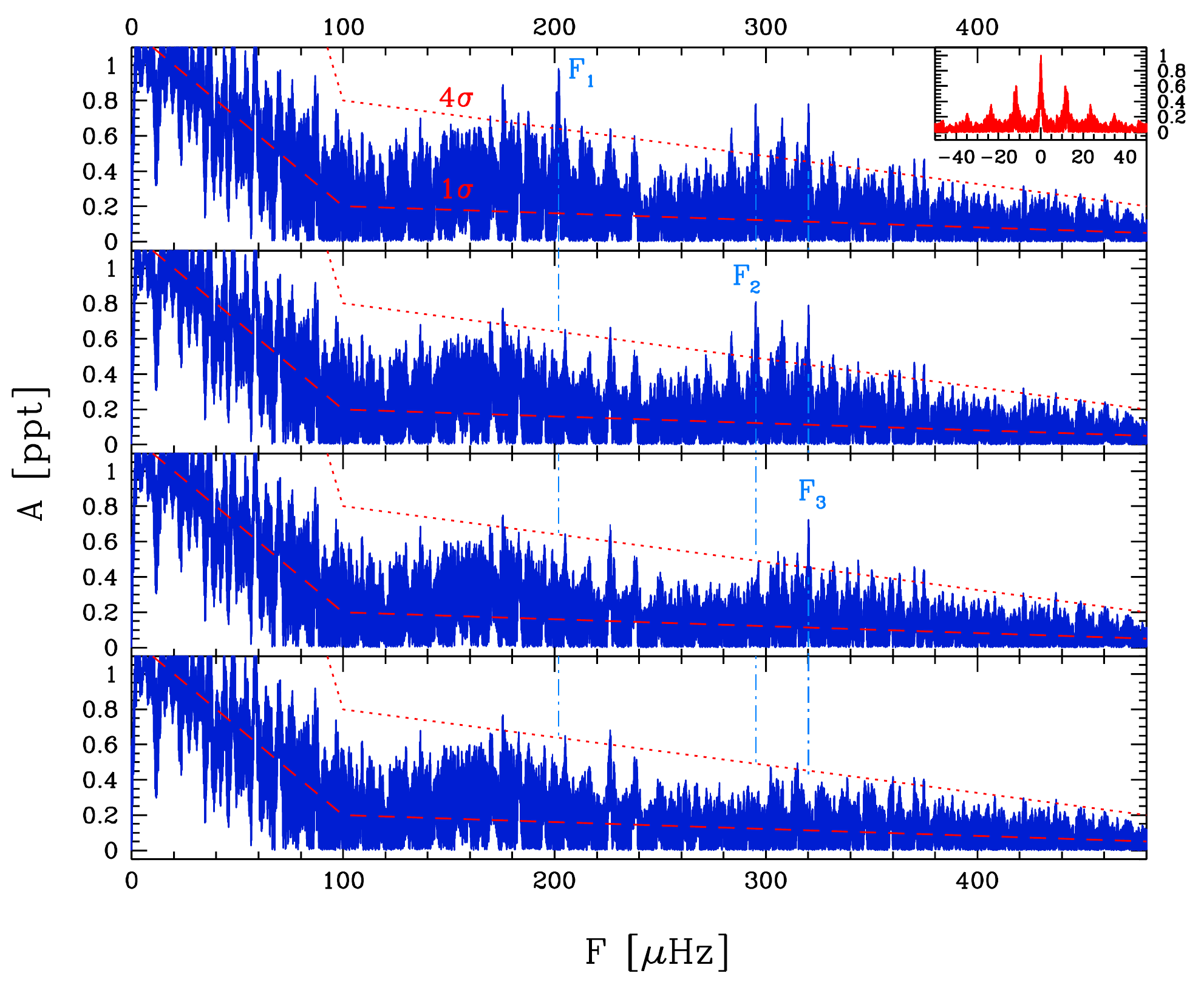}
\caption{$G$-mode amplitude spectrum using our best-quality runs
between 2002 and 2008 (192.8 hours of observations in total).
The upper right panel shows the spectral window (red), while the other panels 
from top to bottom show amplitude spectrum
and residuals after one, two, and three prewhitening steps.
We note an excess of power in two main regions near 180 and 310~\muHz.
After prewhitening, this excess of power is not completely removed near
180~\muHz, suggesting that further low-amplitude frequencies are present 
in that region.}
\label{fig11}
\end{figure}

\section{Summary}

Interpreting the new O-C results shown in Fig.~8 and 9 is more complicated 
than it was ten years ago.
At that time, the very good agreement between the sine-wave component of $f_1$
and $f_2$  strongly supported the presence of a giant planet (SSJ07).
Now, with many more data, this agreement is much more uncertain and the 
presence of V391~Peg~b is weaker and requires confirmation with an 
independent method.
Like in SSJ07, a two-component fit (parabola + sine wave) still gives 
satisfactory results for both $f_1$ and $f_2$, at least up to 2009.
The sinusoidal components of $f_1$ and $f_2$ , however, now differ in period 
and amplitude by $\sim$20\% and $\sim$36\%, respectively.
Starting in phase, after two cycles the O-C sine wave of $f_2$ is antiphased 
with respect to $f_1$.
When we consider all the O--C data from 1999 to 2012, a two-component fit
is in general not satisfactory. 
For $f_1$, we tried to fit the O--C data with a double sine wave,
corresponding to a sinusoidal behavior of \pdot$_1$.
The result is a very poor fit.
However, this solution produces a certain agreement between the sinusoidal 
components of $f_1$ and $f_2$.

The change in sign of the time derivative of the main pulsation period 
near 2009 is an intriguing phenomenon that is difficult to explain. 
Nonlinear interactions between pulsation modes seem the most natural 
explanation, but the $l$=0 identification \citep{silvotti10}, which is 
confirmed by the new data, does not help as we cannot invoke resonant mode 
coupling between the components of a multiplet nor resonance between modes 
linked by linear combinations that we do not see.
The irregular behavior of $f_1$ agrees to a certain extent with recent \kep\
results, which showed that sdB pulsation frequencies are in general less 
stable than previously believed.
The \kep\ results are mostly focused on $g$-modes, but a similar behavior 
seems also relatively common for the $p$-modes.
At least this is suggested by our results.

The $l$=1 identification for $f_2$ \citep{silvotti10} is also
confirmed 
by the new data (or at least $l$ must be $>$0). 
A retrograde mode is detected, although at the limit of our detection 
threshold, and this suggests a stellar rotation period of about 40 days.

Using only the data up to 2009.0, we can improve our previous measurements of
\pdot\ for $f_1$ and $f_2$ and obtain  
\pdot$_1$=(1.34$\pm$0.04)$\times$10$^{-12}$ and 
\pdot$_2$=(1.62$\pm$0.22)$\times$10$^{-12}$.
The order of magnitude of these numbers is in agreement with theoretical 
expectations for evolved models of extreme horizontal branch stars 
\citep{charpinet02}, and their positive sign would normally be interpreted as 
an indicator of a stellar expansion.
At least for $f_1$, however, the change in curvature near 2009 implies that 
these numbers are not simply or directly related to the evolutionary timescales
expected from theory, and the situation is more complicated.

Finally, the new data confirm that V391~Peg is a hybrid pulsator,
showing both $p$- and $g$-modes.
The next opportunity for a more detailed study of this star, and in particular 
for the study of the low-frequency part of its Fourier spectrum, is given by 
the TESS mission, which may observe V391~Peg continuously for 54 days 
in about two years from now.
With a better knowledge of the Fourier spectrum at low frequencies
as well, it 
should be easier to confirm or reject the presence of a planet orbiting 
V391~Peg by measuring the spectroscopic radial velocities of the star.

\begin{acknowledgements}
We thank Elia Leibowitz, who made the data collected at the
Wise Observatory available to us, Christopher D.~J. Savoury for helping us with
the ULTRACAM observations and data reduction, and Wen-Shan Hsiao for 
contributing the Lulin data. 
We also thank Patrick Lenz for providing us with a modified version of 
period04, which facilitated the error estimation from the MC simulations.
%Part of this work was supported by the German \emph{Deut\-sche 
%For\-schungs\-ge\-mein\-schaft, DFG\/} project number Ts~17/2--1.
V.~S.~D. and ULTRACAM are supported by STFC grant ST/J001589/1.
L.~M. was supported by the the Hungarian National Research, Development and 
Innovation Office (NKFIH) grant PD-116175 and the J\'anos Bolyai Research 
Scholarship of the Hungarian Academy of Sciences.
\end{acknowledgements}

% WARNING
%-------------------------------------------------------------------
% Please note that we have included the references to the file aa.dem in
% order to compile it, but we ask you to:
%
% - use BibTeX with the regular commands:
%   \bibliographystyle{aa} % style aa.bst
%   \bibliography{Yourfile} % your references Yourfile.bib
%
% - join the .bib files when you upload your source files
%-------------------------------------------------------------------

\end{document}